\documentclass[10pt,journal,sort & compress]{IEEEtran}
\makeatletter
\def\endthebibliography{%
	\def\@noitemerr{\@latex@warning{Empty `thebibliography' environment}}%
	\endlist
}
\makeatother

\ifCLASSOPTIONcompsoc
\usepackage[noadjust]{cite}
\else
\usepackage[noadjust]{cite}
\fi

\ifCLASSINFOpdf
\usepackage[pdftex]{graphicx}
\graphicspath{{figures/}{figures/results/}}
\DeclareGraphicsExtensions{.pdf,.jpeg,.png}
\else
\usepackage[dvips]{graphicx}
\graphicspath{{figures/}{figures/results/}}
\DeclareGraphicsExtensions{.eps}
\fi

\usepackage{amsmath}
\usepackage{amssymb}		
\usepackage{bm}				
\usepackage{multirow, makecell}
\usepackage{lineno,hyperref}
\usepackage[english]{babel}
\usepackage{tikz,times}
\usepackage{epstopdf}
\usepackage{makecell}
\usepackage{soul,color}
\usepackage{graphicx}
\usepackage{dblfloatfix}
\usepackage[utf8]{inputenc}
\usepackage{multirow}
\usepackage{caption}
\usepackage{lscape}
\usepackage{longtable}
\usepackage{amssymb}
\usepackage{wasysym}
\usepackage{supertabular}
\usepackage{textcomp}
\usetikzlibrary{shapes.gates.logic.US,trees,positioning,arrows}
\usepackage{array}
\newcolumntype{C}{>{\centering\arraybackslash}p{2.5cm}}%

\usepackage{enumitem}

\begin{document}
\title{Security of RPL based 6LoWPAN Networks in the Internet of Things: A Review}

\author{Abhishek~Verma{*}\thanks{* Corresponding Author}, \textit{Student Member, IEEE}, and~Virender~Ranga, \textit{Member, IEEE}
	\IEEEcompsocitemizethanks{\IEEEcompsocthanksitem A. Verma and V. Ranga are with the Department of Computer Engineering, National Institute of Technology Kurukshetra, Haryana, India, 136119.\protect\\
		E-mail: abhiverma866@gmail.com, virender.ranga@nitkkr.ac.in}
}

\IEEEtitleabstractindextext{
	\begin{abstract}
Internet of Things (IoT) is one of the fastest emerging networking paradigms enabling a large number of applications for the benefit of mankind. Advancements in embedded system technology and compressed IPv6 have enabled the support of IP stack in resource constrained heterogeneous smart devices. However, global connectivity and resource constrained characteristics of smart devices have exposed them to different insider and outsider attacks, which put users' security and privacy at risk. Various risks associated with IoT slow down its growth and become an obstruction in the worldwide adoption of its applications. In RFC 6550, the IPv6 Routing Protocol for Low Power and Lossy Network (RPL) is specified by IETF's ROLL working group for facilitating efficient routing in 6LoWPAN networks, while considering its limitations. Due to resource constrained nature of nodes in the IoT, RPL is vulnerable to many attacks that consume the node's resources and degrade the network's performance. In this paper, we present a study on various attacks and their existing defense solutions, particularly to RPL. Open research issues, challenges, and future directions specific to RPL security are also discussed. A taxonomy of RPL attacks, considering the essential attributes like resources, topology, and traffic, is shown for better understanding. In addition, a study of existing cross-layered and RPL specific network layer based defense solutions suggested in the literature is also carried out.
\end{abstract}
	
}

\maketitle

\IEEEdisplaynontitleabstractindextext

\IEEEpeerreviewmaketitle

\section{Introduction}\label{Introduction}
\IEEEPARstart{I}{{NTERNET}} of Things\cite{ashton2009internet} is realized by a large scale deployment of Low power and Lossy Networks (LLNs) which are characterized by communication links that have high packet loss and low throughput \cite{COLAKOVIC2018, winter2012rpl}. LLNs restrict the use of traditional computers and communication technologies due to their strict resource constraints. Also, these networks use resource constrained devices (nodes) that operate on low power, require less energy, have small on-board memory, and low computational capabilities \cite{Musaddiq}. \textcolor{black}{Moreover, characteristics like resource constraints, high packet loss, and low network throughput make state-of-the-art routing protocols like Adhoc On-Demand Distance Vector, Dynamic Source Routing, and Open Shortest Path First unsuitable for LLNs \cite{tripathi2014design, radoiperformance}.} 
\textcolor{black}{To handle this issue, a set of standardized protocols has been developed \cite{IoTEnablingTechnologies, Palattella}. These protocols include IEEE $ 802.15.4 $ PHY/MAC for Physical and Data link layer,  IPv$ 6 $ over Low Power Wireless Personal Area Networks protocol ($ 6 $LoWPAN) for Adaptation layer, Routing Protocol for Low-Power and Lossy Networks protocol (RPL) for Network layer, and Constrained Application Protocol (CoAP) for Application layer.} \textcolor{black}{In transport, layer the standard User Datagram Protocol \cite{rfcUDP} is used.} \textcolor{black}{RPL has been standardized in $ 2012 $ as RFC $ 6550 $ by Routing Over Low power and Lossy networks (ROLL) working group of  Internet Engineering Task Force (IETF) \cite{winter2012rpl}.} 

RPL has been standardized as a network layer protocol for LLNs \cite{Iova}. It is recommended for facilitating efficient routing in LLNs like 6LoWPAN \cite{RPLNutshell}. RPL has gained much popularity in the industry as well as in academia. The reason is its capability to provide efficient routing among resource constrained smart IP enabled IoT nodes, flexibility in adapting to different network topologies, and Quality of service (QoS) support \cite{ Granjal, Palattella, Tomic, LLNMobility}. RPL constructs a Destination Oriented Directed Acyclic Graph (DODAG) from the physical network topology, in which a gateway node is set as a root (destination) of DODAG. All other nodes perform sensing and data routing. RPL uses low energy consuming mechanisms to support self-organization and self-healing for handling frequent node failures \cite{ghaleb2018survey}. \textcolor{black}{It consumes very few resources while providing efficient routing of IPv$ 6 $ packets.} These capabilities of RPL favor its usage in IoT applications that run on LLN infrastructure \cite{kharrufa2019rpl}.  In Section \ref{RPL Concepts and Security Concerns}, a detailed description of RPL is presented. RPL protocol based networks inherit vulnerabilities from its core technologies like IPv$ 6 $ and Wireless Sensor Networks (WSN). Also, Self-organization, self-healing, open nature, and resource constrained characteristics of RPL expose it to various threats that target it for compromising users' security and privacy \cite{RIAHISFAR2018118}. Also, RPL is exposed to external threats from the Internet \cite{MendezPY17, grgic2016security}. Traditional cryptography based security solutions are not suitable for securing RPL based networks (e.g., LLNs). This is because the effectiveness of traditional cryptography based security solutions (e.g. symmetric and asymmetric) relies on the secure distribution of keys. The resource constrained nature of LLNs pose many challenges to key management \cite{el2016performance,ilia2013cryptographic}. Also, if a single legitimate node is compromised, an attacker may gain access to a large pool of pre-loaded keys \cite{bechkit2013highly}. This means, once pre-loaded keys are compromised, all network nodes are also compromised. The challenges related to secure key establishment, storage, distribution, revocation, and replacement in LLNs make traditional cryptography based security solutions unsuitable for LLNs \cite{tsao2015security}. The limitations of LLNs pose a severe threat to RPL security. \textcolor{black}{RPL is vulnerable to various routing attacks, which can be broadly classified into two categories, i.e., attacks inherited from WSN, and RPL specific attacks.} The cryptography based security mechanisms can only prevent RPL from external attacks (i.e., attack performed using a node which is not a part of the existing network) \cite{abomhara2015cyber, WCWC052018005.}. Traditional security mechanisms are also incapable of detecting insider attacks (i.e., attack performed on the devices that are already part of the existing network), which are performed by the compromised nodes of the network \cite{Homoliak}. Thus, from RPL's security point of view, it is crucial to explore the possibilities of developing energy efficient security solutions.

\subsection{Related surveys}
In the literature, some research works particular to RPL, and IoT security are present. Airehrour \textit{et al.} \cite{airehrour2016secure} surveyed various attacks and defense mechanisms specific to RPL. Their study primarily focused on the utilization of trust based defense methods in RPL security. Most of the defense mechanisms, they discussed are used in WSN security and cannot be directly applied to IoT networks. Alaba \textit{et al.} \cite{Alaba2017} presented a detailed review of IoT security issues. However, they did not focus on the RPL protocol. A detailed survey on protocols available for facilitating secure communications in IoT is done by Granjal \textit{et al.} \cite{Granjal}. The authors discussed research challenges for different protocols, including RPL. However, they did not discuss attacks and defense mechanisms specific to RPL. Wallgren \textit{et al.} \cite{wallgren2013routing} did a detailed study on RPL security by implementing some routing attacks and analyzing the network's performance. Also, they suggested the possible mitigation methods of such attacks. However, they only considered WSN based attacks. Mayzaud \textit{et al.} \cite{Mayzaud2016Taxanomy} provided a survey on RPL based attacks and their countermeasures. They proposed a detailed taxonomy of attacks. However, their study did not include the latest proposed attacks and defense solutions. Moreover, the authors did not propose any taxonomy of defense solutions. Pongle \textit{et al.} \cite{pongle2015real} performed a short study on attacks against RPL and $ 6 $LoWPAN layer. The authors only provided a short description of defense solutions and did not propose any suitable taxonomy of attacks and defense solutions specific to RPL. In our opinion, all the mentioned surveys lack effective future research directions and recent proposals. Moreover, these surveys have neglected cross-layered security solutions, which can be utilized for securing RPL as well. The key points that lacked in the previous literature are addressed in our study. The existing surveys related to RPL protocol security are summarized in Table \ref{tab:RelatedComp}.

\subsection{Motivation and contributions}

\textcolor{black}{With an increase in the number of resource constrained devices (LLNs nodes) and their integration with the Internet has led to severe cybersecurity risks.} These risks involve users' security and privacy getting exposed to various threats. Critical applications like healthcare and smart grid, when exposed to such threats, may cause life-threatening incidents to the world population. This motivated us to explore and perform an in-depth analysis of various security issues and their available solutions specific to the RPL protocol.  Since RPL is one of the most popular routing protocols for resource constrained networks hence its security aspect must be studied carefully. In this research paper, we present a comprehensive study of different attacks specific to RPL protocol and their defense solutions suggested in the literature. \textcolor{black}{Our objectives include: ($ 1 $) to propose a taxonomy for classifying different attacks and defense solutions specific to RPL protocol, ($ 2 $) to identify open research issues, and state-of-the-art challenges related to RPL based IoT network security.} The main contributions of this paper are summarized below:
\begin{itemize}
	\item We provide a comprehensive overview of RPL protocol while focusing on its security issues.
	\item We present an extensive survey on RPL specific attacks and their countermeasures present in the recent literature.
	\item We represent the RPL security solutions into two broad categories (i.e., Secure Protocol and Intrusion Detection System) and compare their performance based on different evaluation metrics.
	\item We discuss cross-layered security solutions present in the literature, which can be used to leverage RPL security.
	\item We provide open issues, research challenges, future research directions, and potential areas for future research to promote the contribution of state-of-the-art defense solutions.
	
\end{itemize}

\begin{table*}[!h]
	\centering
	\caption{Comparison with related survey papers}
	\label{tab:RelatedComp}
	\color{black}\begin{tabular}{|p{2cm}|p{3.1cm}|p{4.7cm}|p{3.7cm}|p{2.3cm}|}
		\hline
		\textbf{Related survey} & \textbf{Brief summary} & \textbf{Topics} & \textbf{Scope} & \textbf{Common points with our survey} \\ \hline
		Airehrour \textit{et al.} \cite{airehrour2016secure} & A survey on existing routing protocols and mechanisms to secure routing communications in IoT & Security and  energy consumption in IoT, Routing  protocols, Vulnerabilities to routing, Secure routing protocols, Trust in secure routing & Vulnerabilities in RPL, WSN based defense methods, research challenges & Overview of RPL \\ \hline
		Alaba \textit{et al.} \cite{Alaba2017} & A detailed discussion on the IoT security scenario and analysis of the possible attacks & IoT overview, Classification of IoT, Threats and vulnerabilities,  IoT security taxonomy, Possible  attacks on IoT & State-of-the-art security threats and vulnerabilities, future directions & IoT architecture \\ \hline
		Granjal \textit{et al.} \cite{Granjal} & A detailed survey on protocols available for facilitating secure communications in IoT & IoT communication protocols, Security requirements of IoT, Security of various layers (Physical  (PHY), MAC, network, application layers), Security for routing & IoT communication protocols and their security issues, protocol specific research challenges & Overview of RPL \\ \hline
		Wallgren \textit{et al.} \cite{wallgren2013routing} & A detailed study on RPL security by implementing and analyzing various routing attacks & IoT technologies and IDS, IoT protocols overview, Attacks against  RPL (inherited from WSN), IDS and the  IoT (lightweight heartbeat protocol) & Theoretical impact analysis of attacks inherited from WSN, defense mechanism to detect selective forwarding attack. & Overview of RPL, Attacks against RPL (inherited from WSN) \\ \hline
		Mayzaud \textit{et al.} \cite{Mayzaud2016Taxanomy} & A survey on RPL based attacks and their countermeasures & ·RPL concepts  and security concerns, Attacks against RPL protocol, Exploitation for risk management & Attacks against RPL protocol, risk management for RPL security & RPL overview and attacks \\ \hline
		Pongle \textit{et al.} \cite{pongle2015real} & A short study on attacks against RPL and 6LoWPAN layer & Overview of RPL and 6LoWPAN, Attacks on RPL topology, IoT and IDS, Attacks on 6LoPWAN layer & RPL and 6LoWPAN adaptation layer security & Overview of  RPL, Attacks on RPL  topology \\ \hline
		Verma \textit{et al.} (Our survey) & A detailed survey on security of RPL based 6LoWPAN Networks & IoT architectures, Overview of RPL, RPL specific  attacks, Taxonomy of RPL attack defense mechanisms, Secure  protocol approaches, IDS, Cross-layered security solutions & RPL specific attacks and their countermeasures , research challenges, future directions & - \\ \hline
	\end{tabular}
\end{table*}

\begin{figure}[!h]
	\centering
	\includegraphics[scale=.8]{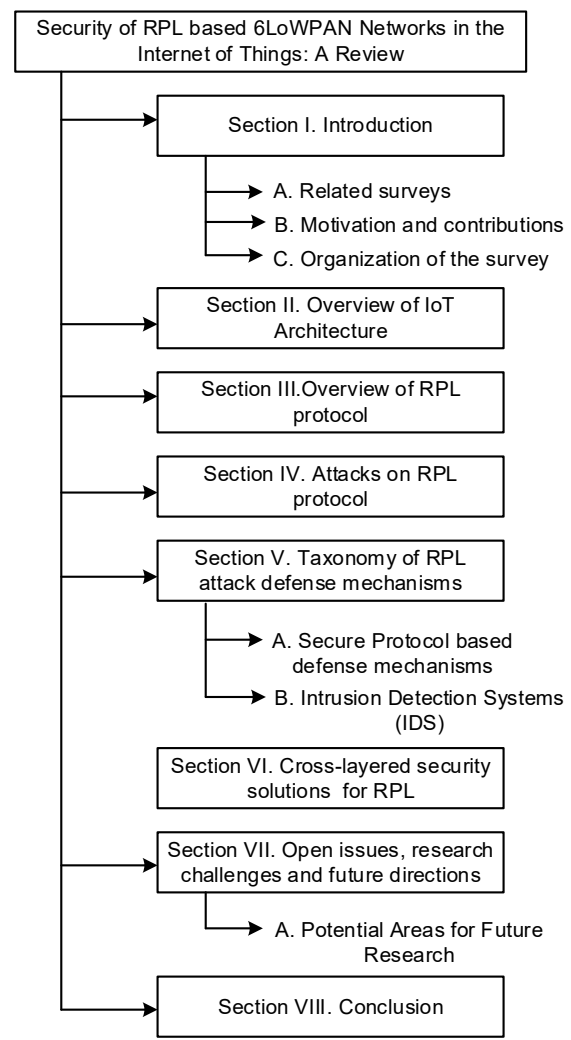}
	\caption{Organization of the survey}
	\label{org}
\end{figure}

\subsection{Organization of the survey}

The rest of this paper is consequently organized as follows. Section \ref{Basics of Internet of Things} describes the overview of IoT architectures.
Section \ref{RPL Concepts and Security Concerns} presents a brief overview of the RPL protocol. Section \ref{Taxonomy of Attacks on RPL protocol} presents a taxonomy of attacks specific to the RPL protocol. In Section \ref{Taxonomy of RPL attack defense mechanisms}, the proposed taxonomy related to different defense solutions against RPL attacks present in the literature is discussed. \textcolor{black}{In Section \ref{crosslayered}, cross-layered security solutions specific to RPL protocol security are discussed.} Open issues, research challenges, and future directions are addressed in Section \ref{Open issues and research challenges}. Finally, the paper is concluded in Section \ref{Conclusion}. The list of abbreviations and definitions used throughout the paper are presented in Table \ref{LOA}. The organization of the survey is illustrated in Fig. \ref{org}.

\begin{table}[]
	\centering
	\caption{List of Abbreviations}
	\label{LOA}
	\begin{tabular}{ll}
		\textbf{Abbreviation} & \textbf{Stands For} \\
		IoT & Internet of Things \\
		IP & Internet Protocol \\
		IPv6 & Internet Protocol version 6 \\
		RPL & IPv6 Routing Protocol for Low Power and Lossy Network \\
		IETF & Internet Engineering Task Force \\
		ROLL & Routing Over Low power and Lossy Networks \\
		6LoWPAN & IPv6 over Low Power Wireless Personal Area Networks \\
		LLNs & Low power and Lossy Networks \\
		CoAP & Constrained Application Protocol \\
		UDP & User Datagram Protocol \\
		QoS & Quality of Service \\
		DODAG & Destination Oriented Directed Acyclic Graph \\
		WSN & Wireless Sensor Networks \\
		ML & Machine Learning \\
		WLAN & Wireless Local Area Network \\
		WPAN & Wireless Personal Area Networks \\
		LoWPAN & Low-Power Wireless Personal Area Networks \\
		GSM & Global System for Mobile Communications \\
		NFC & Near-field communication \\
		LTE & Long-Term Evolution \\
		OF & Objective Function \\
		ETX & Expected Transmission Count \\
		MRHOF & Minimum Rank with Hysteresis Objective Function \\
		OF0 & Objective Function Zero \\
		OF-EC & OF based on combined metrics using Fuzzy Logic \\
		DIO & DODAG Information Object \\
		DIS & DODAG Information Solicitation \\
		DAO & Destination Advertisement Object \\
		DAO-ACK & Destination Advertisement Object Acknowledgment \\
		DoS & Denial-of-Service \\
		PDR & Packet Delivery Ratio \\
		6BR & 6LoWPAN Border Router \\
		IDS & Intrusion Detection System \\
		VERA & Version Number and Rank Authentication \\
		TRAIL & Trust Anchor Interconnection Loop \\
		SRPL & Secure-RPL \\
		TCA & Trusted Computing Architecture \\
		TPM & Trusted Platform Module \\
		MRTS & Metric based RPL Trustworthiness Scheme \\
		ERNT & Extended RPL Node Trustworthiness \\
		TIDS & Trust based Security System \\
		TOF & Trust Objective Function \\
		TRU & Trust Information \\
		AT & Adaptive Threshold \\
		DT & Dynamic Threshold \\
		FAM & Frequency Agility Manager \\
		LR & Logistic Regression \\
		MLP & Multi-layer Perceptron \\
		NB & Naive Bayes \\
		RF & Random Forest \\
		SVM & Support Vector Machine \\
		SOMIDS & Self Organizing Map Intrusion Detection System \\
		SOM & Self Organizing Map \\
		RSSI & Received Signal Strength Indicator \\
		TN & True Negative \\
		FP & False Positive \\
		FN & False Negative \\
		TPR & True Positive Rate \\
		FPR & False Positive Rate \\
		SPRT & Sequential Probability Ratio Test \\
		InDRes & Intrusion detection and response system for IoT \\
		FSM & Finite State Machine \\
		EFSM & Extended Finite State Machine \\
		SBIDS & Sink-based Intrusion Detection System \\
		NCR & Node's current rank \\
		NPR & Node's parent rank \\
		NPVR & Node's previous rank \\
		RIDES & Robust Intrusion Detection System \\
		CUSUM & Cumulative Sum Control charts \\
		OPFC & Unsupervised Optimum-Path Forest Clustering \\
		NAC & Network Access Control \\
		ETA & Encrypted Traffic Analytics \\
		6TiSCH & IPv6 over the TSCH mode of IEEE 802.15.4e \\
		TSCH & Time-Slotted Channel Hopping
	\end{tabular}
\end{table}

\section{Overview of IoT Architecture}\label{Basics of Internet of Things}

\begin{figure}[!h]
	\centering
	\includegraphics[scale=.63]{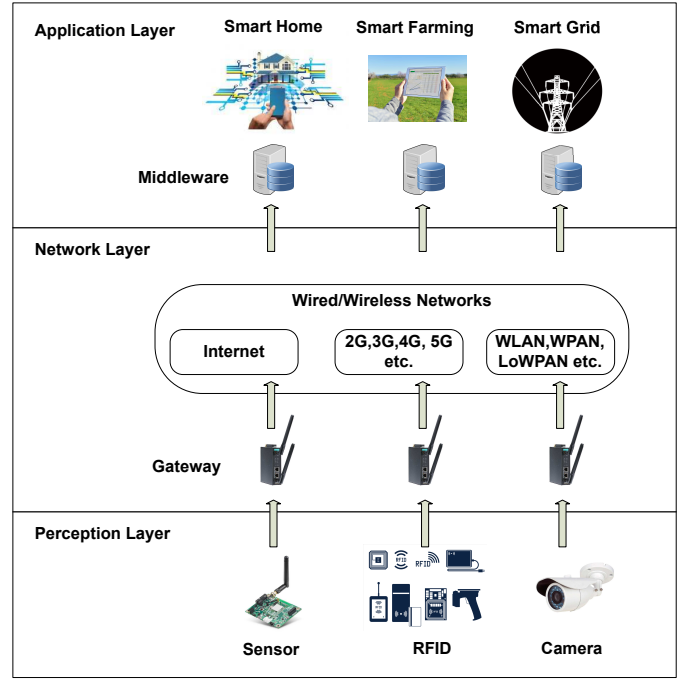}
	\caption{Three-layer architecture of IoT}
	\label{3LayeredArchitecture}
\end{figure}

Various architectures applicable to IoT have been proposed in the literature. \textcolor{black}{Most popular architectures include middleware based, service-oriented based, three-layer and five-layer based \cite{sethi2017internet}.} Any standard IoT architecture is not yet recognized in the literature. However, the most commonly referred IoT architecture is three-layer-based architecture \cite{Abdmeziem2016}, which is shown in Fig. \ref{3LayeredArchitecture}. \textcolor{black}{It gains popularity because of simple nature and abstract representation of IoT that makes the implementation of applications easier. It comprises three layers, namely perception, network, and application. These layers are highlighted below.}

\subsubsection{Perception Layer}
The perception layer is the lowest layer in three-layer IoT architecture. The main purpose of the perception layer is to collect data from the physical environment (temperature, pressure, humidity, etc.) of IoT devices. The process of perception is supported by prominent sensing technology like WSN. Besides, this layer is responsible for converting analog input to digital form and making sensed data suitable for transmission.

\subsubsection{Network Layer}
The network layer is dedicated to the processing of sensed data and performing secure data transmission between the perception and application layer. It uses various wired and wireless networking technologies like WLAN, WPAN, LoWPAN, and GSM. It integrates various transmission technologies like NFC, LTE, and Bluetooth. It promises unique addressing and routing of sensed data from a large number of devices, which are a part of the IoT network. $ 6 $LoWPAN is standardized for achieving unique addressing through IPv$ 6 $ networking.

\subsubsection{Application Layer}
The main purpose of the application layer is to provide personalized services or interface (front end) to the IoT application users. It uses processed data from the network layer and delivers it as per the user's need. \textcolor{black}{It fills the gap between users and IoT applications. The application layer provides tools to the application developers in order to realize IoT insights.} It specifies various applications in which IoT can be exploited, e.g., smart homes, smart power grid, industrial monitoring, surveillance systems, healthcare monitoring, and logistics management \cite{ IoTCore1,MIORANDI20121497, sethi2017internet}

\begin{figure}[!h]
	\centering
	\includegraphics[scale=.63]{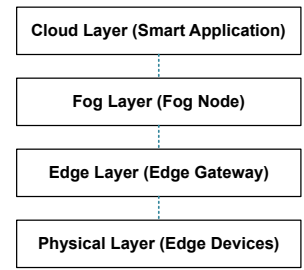}
	\caption{\textcolor{black}{Fog computing based four-layer IoT architecture}}
	\label{FOA}
\end{figure}

\textcolor{black}{Currently, IoT devices generate a large volume of data that needs to be processed at cloud servers for various purposes like business insights and security monitoring.} However, the rate at which data is generated by IoT devices, requires good bandwidth connections for data transmission to the cloud servers. The limited bandwidth connections cause a significant delay in data transmission and processing, which affects the overall performance of smart applications. Three-Layer based architecture is not capable enough to solve such issues \cite{sethi2017internet}. To address these issues,  Edge and Fog computing paradigms \cite{dastjerdi2016fog} are emerged as possible solutions and are being used nowadays. The four-layered Fog computing based IoT architecture is shown in Fig. \ref{FOA}. The physical layer includes IoT devices or edge devices which sense and send data to edge gateways. \textcolor{black}{The edge layer consists of edge gateways (border routers), which either perform real-time data preprocessing at source/on-premises or forward the received data to the fog node. The fog layer consists of powerful servers that collect data from edge gateways and perform the task of data preprocessing, and transmission to the cloud servers.} At the cloud layer, smart applications are deployed, which perform critical tasks like business insights and security monitoring. \textcolor{black}{Fog nodes can process and act on a large volume of data, reduce bandwidth and latency, and can perform security monitoring.} Whereas edge nodes can apply local security policies and make real-time decisions locally to control and monitor many IoT devices at a time. With fog and edge layer, the security of IoT application and involved protocols can be improved significantly \cite{mukherjee2017security, yi2015security, prabavathy2018design}. 

\section{Overview of RPL protocol}\label{RPL Concepts and Security Concerns}
RPL is IPv6 based Distance Vector and Source Routing protocol that specifies how to build a DODAG using a \textit{Objective Function (OF)}, set of metrics and constraints. In RPL, the IoT devices are interconnected using mesh and tree topology in order to build a DODAG graph starting from a root (sink or gateway) node that acts as an interface between LLN nodes and the Internet. A network may contain more than one DODAG, which collectively form an RPL Instance. In a network, more than one RPL Instance can run in parallel, and every RPL Instance is identified by a unique \textit{RPLInstanceID}. An RPL node can belong to only one DODAG of every RPL Instance running in the network. Each node in DODAG is assigned a rank (16-bit value), which represents ``the node's individual position relative to other nodes with respect to a DODAG root" \cite{winter2012rpl}. The rank stringently increases in DODAG's downward direction (root to leaves) and decreases in the upward direction (leaf nodes to root). \textcolor{black}{The rank concept is used: (1) to detect and avoid routing loops, (2) to build parent-child relationship, (3) to provide a mechanism for nodes to differentiate between parent and siblings, and (4) to enable nodes to store a list of preferred parents and siblings which can be used in case a node loses its link with the parent node.} DODAG is built during the network topology setup phase, where each node uses RPL control messages to find the optimal set of parents towards the root and link itself with the preferred parent, i.e., parent on the most optimal path. The selection of preferred parent is based on a \textit{OF} that defines how to compute a rank based on routing metrics while considering routing constraints and optimization objectives. RPL may use different \textit{OF} \cite{lamaazi2020comprehensive} which includes ETX Objective function \cite{ETXOF}, \textit{Minimum Rank with Hysteresis Objective Function (MRHOF)} \cite{MRHOF}, \textit{Objective Function Zero} \textit{(OF0)} \cite{OF0}, and objective function based on combined metrics using fuzzy logic \textit{(OF-EC) \cite{lamaazi2018ec}}. RPL control messages include DODAG Information Object (DIO), DODAG Information Solicitation (DIS), Destination Advertisement Object (DAO), and Destination Advertisement Object Acknowledgment (DAO-ACK). \textcolor{black}{RPL uses an adaptive timer mechanism called as ``Trickle timer" in order to limit the control traffic in the network \cite{levis2011trickle}.} 

\section{Attacks on RPL protocol}\label{Taxonomy of Attacks on RPL protocol}
\begin{figure*}[!h]
	\centering
	\includegraphics[scale=0.6]{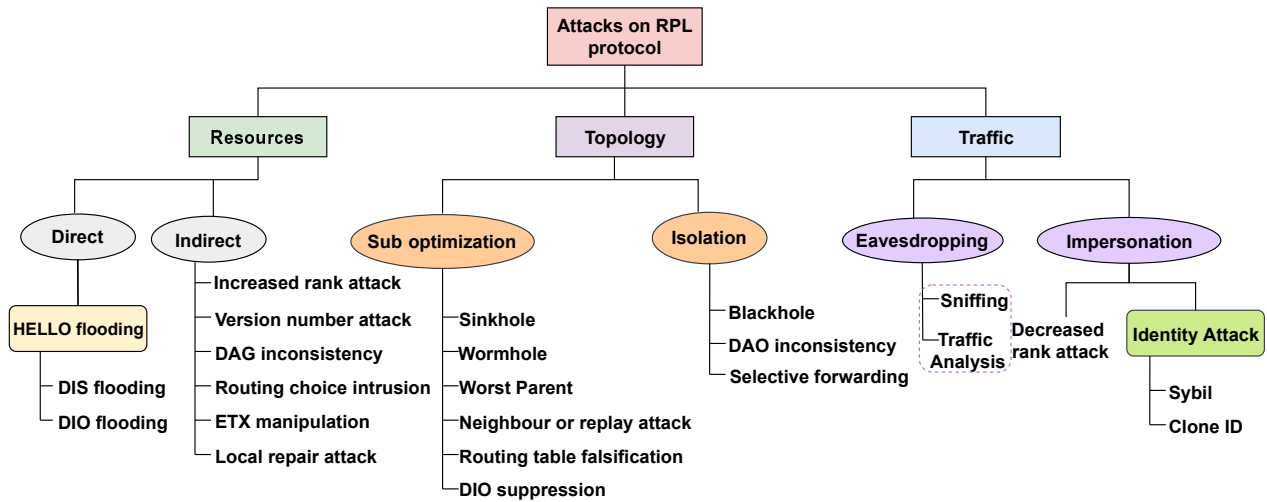}
	\caption{Detailed taxonomy of attacks specific to RPL protocol}
	\label{RPLTaxonomy}
\end{figure*}

RPL protocol is susceptible to a wide range of insider and outsider attacks. These attacks are difficult to detect and mitigate because of the vulnerable nature of nodes and wireless network, easily tamperable nature of nodes, mobility of nodes, and resource constraints. Many authors have proposed various security mechanisms specific to RPL, which include control message encryption and security modes \cite{Perazzo2017a}. \textcolor{black}{However, most of the RPL implementations do not consider the security measures due to incomplete specification of mechanisms, and implementation overheads \cite{KAMGUEU201810}.} These security mechanisms are effective in defending against outsider attacks. However, they fail in case of insider attacks. \textcolor{black}{An insider attacker may bypass the applied RPL security mechanisms and disrupt network functionality.} \textcolor{black}{A taxonomy of attacks, is shown in Fig. \ref{RPLTaxonomy}, where attacks are classified on the basis of their primary target.} We have extended the taxonomy presented in \cite{Mayzaud2016Taxanomy} by adding recently proposed attacks, and categorizing some similar kind of attacks for better understanding. RPL control messages can be illegitimately manipulated to disrupt routing operations. Similarly, fault tolerance mechanisms can be exploited to target network resources by performing a Denial of Services attack (DoS). In this section, attacks specific to the RPL protocol are listed and briefly discussed.

\textit{Rank attacks}: The rank field or rules can be exploited for performing various rank based attacks \cite{xie2010routing}. In RPL, there is no specific mechanism to monitor the integrity of control messages and routing metric values received from the parent node. In fact, a child node receives all the routing information through control messages without verifying its parent trustworthiness. Thus, if the parent node is malicious, the child node still believes that all the information coming from its parent is genuine. Hence, this condition may lead to the formation of unoptimized routes and show poor network performance. \textcolor{black}{An attacker node performs the Rank attack by illegitimately changing its rank value, thus, attracting neighbor nodes to select it as their parent, assuming that the malicious node leads to the root node in the shortest path cost.} Different variants of Rank attack have been proposed in the literature by the researchers, which include increased rank, decreased rank, worst parent attacks.

\textit{Neighbor or replay attack}: In neighbor attack \cite{Le2013}, an attacker node duplicates and multicast all DIO messages received from its parent. \textcolor{black}{In such a case, all the neighbor nodes which receive replayed DIO messages may think that the message is received from a new neighbor.} \textcolor{black}{Further, if the replayed DIO message contains favorable routing information like rank, the victim neighbor node may add out of range node as its preferred parent. Another variant of this attack is proposed in \cite{Mayzaud2016Taxanomy} and termed as DIO replay attack.} In this variant, an attacker nodes multicast the outdated DIO messages containing old routing information. \textcolor{black}{This attack forces a victim node to follow the stale and unoptimized paths.}  

\textit{DAO attacks}: An adversary can exploit the storing mode of the RPL protocol. It can manipulate the DAO messages to perform DAO related attacks. These types of attacks include DAO inconsistency and routing table falsification. \textcolor{black}{Both of these are highlighted below.}      

\textit{DAO inconsistency}: \textcolor{black}{RPL uses some flags which are carried out in IPv6 hop-by-hop option to manage important topological mechanisms. Down `O' flag represents the expected direction of packet, Rank-Error `R' flag indicates rank error in topology, and Forwarding-Error `F' flag represents that the node is not capable of forwarding packet to the set destination \cite{winter2012rpl}. DAO inconsistency is reported by a node when its child node is unable to forward the data to a specified destination, due to unavailability of a route that is learned from fake DAO message (DAO with fake routing information) during topology creation.} The attacker exploits this mechanism to perform an attack by setting `F' flag to $ 1 $ in the packets and sending it back to its parent. This forces the parent node to discard legitimate available downward routes. DAO inconsistency attack leads to an increase in end-to-end delay, unoptimized topology, and isolation of nodes.

\textit{Routing table falsification}: \textcolor{black}{Mayzaud \textit{et al.} \cite{Mayzaud2016Taxanomy} proposed a methodology to perform the attacks that lead the nodes to learn fake routes which do not exist. Such attacks can create unoptimized topology due to increased end-to-end packet delay and decreased packet delivery ratio (\textit{PDR})}. \textcolor{black}{An attacker may perform the attack by forging the routing information contained in DAO messages, which forces the legitimate nodes to build false downward routes, i.e., non-existing routes.} Thus, when legitimate nodes try to forward the data to non-existing nodes, this situation leads to DAO inconsistency, unnecessary packet delay, and increased control overhead. In a variant of routing table falsification attack termed as routing table overload, the attacker forges a DAO message with false routing information and sends it to the parent node. \textcolor{black}{It leads to the victim node's routing table buffer getting full.} \textcolor{black}{Thus, further creation of legitimate optimized routes is entirely blocked.}     

\textit{Routing choice intrusion}: Zhang \textit{et al.} \cite{zhang2015intrusion} proposed a new internal routing attack known as Routing choice intrusion. \textcolor{black}{The main idea is to learn the current routing conditions used by the nodes for choosing optimal paths.} Then capturing the DIO messages, and later multicast the forged DIO messages by its legitimate identity. \textcolor{black}{This attack requires a node to be reprogrammed in such a manner that it ignores the internal misbehavior detection and operates normally, thus, makes it hard to be detected.} This attack may involve one or more compromised nodes. Routing choice intrusion attack leads to an increase in end-to-end delay, routing loops, energy consumption, and creation of unoptimized paths. 

\textit{DIS attack}: In DIS attack, an attacker node periodically sends DIS messages to neighbors within its transmission range. \textcolor{black}{In return, the victim node resets its trickle timer and replies with DIO messages (RPL specific mechanism for allowing new nodes to join DODAG) \cite{vermaETT,TENCON2019}.} \textcolor{black}{This attack can be performed either by sending unicast DIS messages to a single node or by multicasting DIS messages in order to target multiple nodes at a time.} DIS attacks can be termed as flooding attack as it involves the flooding of DIS messages in the network \cite{Mayzaud2016Taxanomy}. It leads to an increase in control packet overhead, node energy exhaustion, and routing disruption.   

\textit{Version number attack}: \textcolor{black}{In RPL, only border router ($ 6 $BR) is responsible for initiating the propagation and updation (increase) of version number \cite{winter2012rpl}.} {Whenever a border router or gateway (6BR) needs to rebuild the whole DODAG, it initiates a global repair process by incrementing the version number value present in the version number field of DIO message and sends it to child nodes.} Upon receiving a DIO with a different version number than it has, the child node starts the process for updating its routing state (preferred parent, preferred parent, and links) by resetting its trickle timer. This process iterates until all the nodes update their routing state. RPL defines no mechanism to prevent nodes (other than $ 6 $BR) from illegitimate modification of version number \cite{Dvir2011,landsmann2013topology, Mayzaud2016Version}. \textcolor{black}{Hence, an attacker can modify the version number field of the DIO message and forwards it to the neighbors.} This leads to the unnecessary rebuilding of complete DODAG. It results in an increase in control packet overhead, end-to-end delay, rank inconsistencies, routing loops, and energy consumption. 

\textit{Local repair attack}: In RPL, a local repair mechanism is triggered by a node after it loses the link with its preferred parent \cite{winter2012rpl}. \textcolor{black}{A node can initiate a local repair mechanism either by changing the value in the DODAG ID field of DIO or by updating its rank to infinite and multicast the DIO to all its neighbors.} Both the methods force neighbor nodes to search for a new preferred parent. Local repair enables an RPL network to converge once again in minimum time. This mechanism is supposed to be called only when a node \textcolor{black}{does} not have any connection with its parent. \textcolor{black}{However, an attacker may deliberately use both the methods to trigger unnecessary local repairs even if it is still connected to its parent \cite{Le2012, Le2011, Le2016}.}  This is possible because RPL \textcolor{black}{does} not define any method that can be used by a node to verify the authenticity of local repair initiated by their neighbor nodes \cite{tsao2015security}. \textcolor{black}{Wherever a local repair is triggered, the network topology is forced to be restructured.} \textcolor{black}{This leads to an increase in energy consumption of victim nodes as well as disruption of the routing process.}       

\textit{DODAG inconsistency}: RPL specifies the data path validation method to detect and repair rank 
related inconsistencies (loops) in DODAG. RPL uses different flags of RPL IPv6 header options of multi-hop data packets \cite{hui2012routing} for tracking inconsistencies (routing loops) in the network. \textcolor{black}{As per \cite{winter2012rpl}, DODAG is inconsistent if the direction flag of the data packet represented by the `O’  does not follow the strict rank relation with the node that has sent/forwarded the packet.} When such a situation is encountered, the `R’ flag is used to perform topology repair, i.e., `R' flag is set to $ 1 $ by the node which encountered forwarding error, and the packet is forwarded. Further, when another node receives a packet with `R' flag set (detects inconsistency), it discards the packet and resets its trickle timer to perform local repair \cite{levis2011trickle}. An attacker can exploit these flags to perform various attacks that are collectively termed as a DODAG inconsistency attack, which includes Direct and Forced blackhole attack \cite{Sehgal2014, Mayzaud2015}.

\begin{figure}[!h]
	\centering
	\includegraphics[width=.43\textwidth]{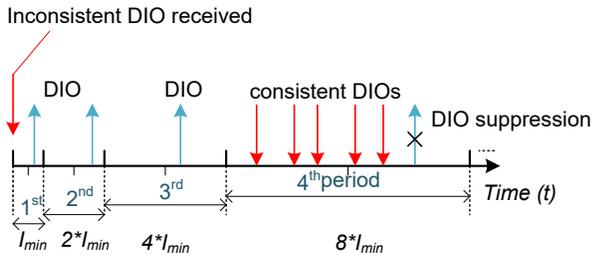}
	\caption{DIO suppression with suppression threshold ($ k=5 $)\cite{Perazzo2017}}
	\label{DIOSuppression}
\end{figure}

\textit{DIO suppression}: In \cite{Perazzo2017}, a novel attack against RPL protocol was proposed and termed as a DIO suppression attack. The idea behind this attack is to suppress the transmission of fresh DIO control messages required by the IoT nodes for exploring new optimized routing paths and removal of stale paths. \textcolor{black}{This leads to the creation of unoptimized routes, which further leads to a partition problem in the network. An attacker only needs to sniff DIO message from any legitimate node and then, multicast that message for at least $ k $ times (suppression threshold) periodically.} This makes victim node believe that the consistent DIOs \cite{levis2011trickle} are received from its parent node irrespective of any legitimate change in network's current state. Thus, there won't be any change in victim's current state, i.e., preferred parent set, parent, and relative distance from the root. In Fig. \ref{DIOSuppression}, $ I_{min} $ represents the starting time period set by trickle algorithm, which is doubled every time $ k $ consistent DIO's are received. $ I_{min} $ is initiated again when DIO's less than $ k  $ are received or when any inconsistent DIO is received.           

\textit{ETX manipulation}: In RPL, the Expected transmission count (ETX) objective function uses the ETX parameter as a metric for selecting the optimal routing path between two nodes. \textcolor{black}{RPL follows a simple thumb rule, i.e., the ETX value of any parent node must be lower than that of a child node. This rule must be followed throughout the network.} An attacker exploits this rule by deliberately manipulating nodes ETX value in order to gain a better position in the network \cite{Shreenivas2017}. This allows the attacker to attract a large part of network traffic and then launch other attacks like Blackhole and Grayhole attacks.

Table \ref{table:COAIP} presents a classification of attacks based on their type (insider or outsider), prerequisites, and their impact on the network's performance. RPL is also vulnerable to attacks inherited from WSN. These attacks include HELLO flood or DIO flood, Sinkhole, Wormhole \cite{deshmukh2019real}, Blackhole, Selective forwarding, Sybil, Clone ID, etc. These attacks disrupt the network's performance drastically, which decreases the network's lifetime. Since many surveys are already available in the literature that present WSN based attacks hence we do not discuss them in this paper \cite{karlof2003secure, bhushan2018recent}.

\begin{table*}
	\centering
	\caption{\textcolor{black}{Classification of attacks on RPL and their impact on network's performance}}
	\label{table:COAIP}
	\color{black}\begin{supertabular}[l]{|p{1.6cm}|p{1.3cm}|p{1.5cm}|p{5.8cm}|p{5.8cm}|}
		\hline
		\textbf{Attack}               & \textbf{Type}    & \textbf{Prerequisites}       & \textbf{Description}                                                                                                                                                                                & \textbf{Impact on network performance}                                                                                                                                                                                \\ \hline

		Rank                          & Insider          & -                            & Rank field and strict rank rules are exploited.                                               & Generates routing loops. Increases end-to-end delay, \textit{PDR}, control packet overhead, congestion, and energy consumption. Introduces unoptimized routes.  \\ \hline
		Neighbor
		/replay              & Insider          & -                            & Attacker node eavesdrops the DIO messages of legitimate neighbors and later send it to its neighbors & Increases packet loss (low \textit{PDR}), disrupted routes, network congestion, and unwanted interference.                                                                                                     \\ \hline
		DAO inconsistency      & Insider          & Storing mode, Option Header  & DAO loop recovery mechanism is exploited by the attacker.                                                                                                                                           & Increases end-to-end delay. Leads to unoptimized topology and isolation of nodes.                                                                                                                                      \\ \hline
		Routing table falsification   & Insider          & Storing mode, Option Header  & Attacker overloads the routing table of legitimate nodes with false routing information.                                                                                             & Routing table buffer of victim nodes gets filled, which further blocks the building of legitimate optimized routes.                                                                                    \\ \hline
		Routing choice intrusion      & Insider          & -                            & Attacker node learns the current routing rules. Then, it captures real DIO messages and multicast the forged DIO messages.                                                                            & Increases end-to-end delay and energy consumption. Generates routing loops and introduces unoptimized paths.                                                                                                                                  \\ \hline
		DIS                           & Insider
		/Outsider & -                            & Legitimate nodes are flooded with DIS messages, which forces them to reset their trickle timer and reply with new DIO messages.                                                                                      & Increases control packet overhead and energy consumption, and causes routing disruption.                                                                                                                            \\ \hline
		Version number                & Insider          & -                            & Attacker node deliberately increments the version number, which triggers global repair of the network.                                                                                     & Increases control packet overhead, end-to-end delay, and energy consumption. Introduces rank inconsistencies and routing loops.                                                                                                    \\ \hline
		Local repair                  & Insider          & -                            & Local repair mechanism is exploited, i.e., by changing the rank value to infinite or changing DODAG ID value to trigger unnecessary local repairs.                                                    & Disrupts the routing process and increases energy consumption.                                                                                                                                                              \\ \hline
		Direct DODAG inconsistency    & Insider
		/Outsider & Option  Header               & Local repair mechanism is exploited, i.e., attacker multicast the packets after setting `O' and `R' flags.          & Traffic congestion. Increases packet loss ratio, control packet overhead and energy consumption.                                                                                                                          \\ \hline
		Forced blackhole              & Insider          & Option  Header               & Attacker node sets `O' and `R' flags of received data packets and forwards them to its neighbors.                   & Increases control packet overhead and energy consumption. Decreases \textit{PDR}.                                                                                                                        \\ \hline
		DIO suppression               & Insider
		/Outsider & -                            & Previously eavesdropped DIO messages are sent, which leads to suppression of new DIO transmission.                                                                                       & Introduces unoptimized routing paths, which leads to network partition.                                                                                                                                                  \\ \hline
		ETX manipulation              & Insider          & ETX objective function       & Manipulation of ETX value in order to gain a better position in the network and attract network traffic.                                                                                                 & Introduces unoptimized routing paths.                                                                                                                                                                                            \\ \hline
		HELLO/DIO flood               & Insider
		/Outsider & -                            & DIO messages with favorable routing metrics are multicast with strong signal strength.                                                                                                              & Leads to network congestion and saturation of RPL nodes. Increases packet loss ratio and control packet overhead.                                                                                                    \\ \hline
		Sinkhole                      & Insider          & -                            & Malicious node decreases its rank in order to become the preferred parent of its neighbors.                                                                                                  & Degrades the overall network performance due to unoptimized routes.                                                                                                                                              \\ \hline
		Blackhole                     & Insider          & -                            & Malicious node drops all the packets it receives from its children nodes.                                                                                                                             & Decreases \textit{PDR}, increases end-to-end delay, unstabilizes topology.                                                                                                                                  \\ \hline
		Selective forwarding/grayhole & Insider          & -                            & Malicious node selectively drops packets, i.e., forwards control packets and drops data packets.                                                                                         & Negatively affects topology construction, which leads to disrupted routing. Decreases \textit{PDR}.                                                                                                          \\ \hline
		Wormhole                      & Insider          & Minimum two malicious nodes. & Two or more nodes create a high bandwidth tunnel between them in order to transmit data in long range.                                                                                                   & Creates unoptimized paths.                                                                                                                                                                                            \\ \hline
		Sybil                         & Insider          & -                            & Single node posses multiple logical identity.                                                                                                                                                       & Overcomes voting schemes, compromises transmission routes by taking control of network.                                                                                                                               \\ \hline
		Clone ID                      & Insider
		/Outsider & -                            & Single logical identity is copied to multiple nodes.                                                                                                                                                 & Compromises transmission routes by taking control of the network, eavesdrop on transmission links.                                                                                                                         \\ \hline
		Jamming                       & Outsider         & -                            & Attacker transmit with high power radio signals to introduce heavy interference.   & Decreases \textit{PDR} and increases energy consumption.                                                                                                \\ \hline
		Sniffing                      & Insider
		/Outsider & -                            & Network traffic is eavesdropped for obtaining routing information from packets.                                                                                                                     & Introduces privacy concerns.                                                                                                                                                  \\ \hline
		Traffic analysis              & Insider
		/Outsider & -                            & Radio transmissions are eavesdropped to analyze traffic patterns for obtaining routing/topology information.                                                                                      & Introduces privacy concerns.\\ \hline
	\end{supertabular}
\end{table*}

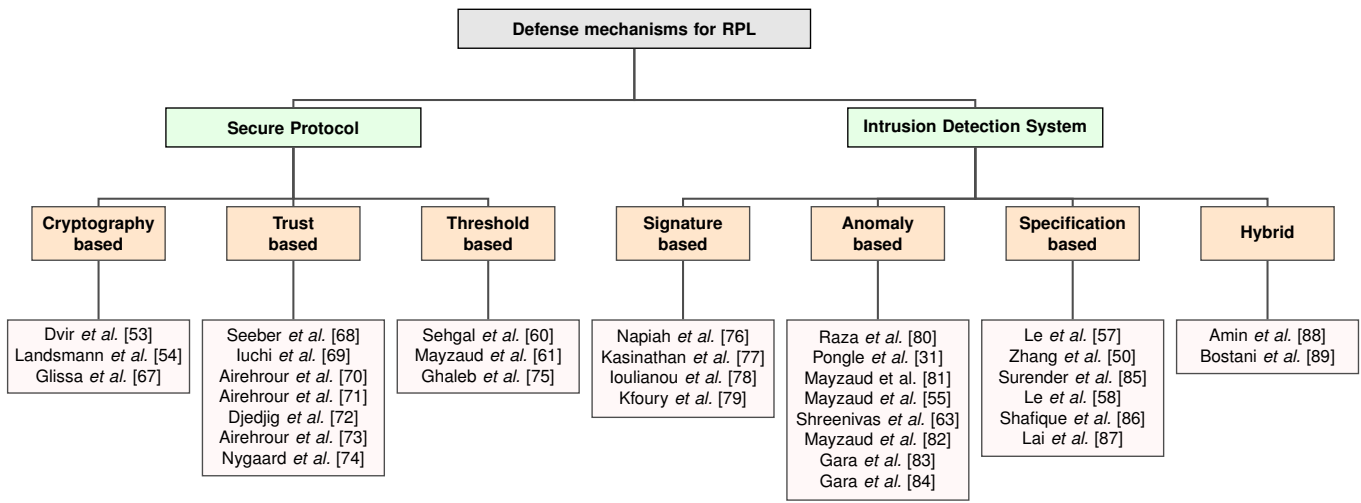
\begin{figure*}[!h]
		\centering
        \resizebox{1\linewidth}{!}{
		\begin{tikzpicture}[
		label distance=3mm,
		every label/.style={black},
		root/.style={rectangle,thick,draw,fill=black!10,text width=7cm,
			text centered,font=\sffamily,anchor=north, minimum height=0.8cm},
		l1/.style={rectangle,thick,draw,fill=green!10,text width=5cm,
			text centered,font=\sffamily,anchor=north,},
		l2/.style={rectangle,thick,draw,fill=orange!20,text width=2.5cm,
			text centered,font=\sffamily,anchor=north, minimum height=1.1cm},
		leaf1/.style={rectangle,thick,draw,fill=red!3,text width=3.5cm,
			text centered,font=\sffamily,anchor=north, minimum height=1.5cm},
		leaf2/.style={rectangle,thick,draw,fill=red!3,text width=3.5cm,
			text centered,font=\sffamily,anchor=north, minimum height=3.2cm},
		leaf3/.style={rectangle,thick,draw,fill=red!3,text width=3.5cm,
			text centered,font=\sffamily,anchor=north, minimum height=1.5cm},
		leaf4/.style={rectangle,thick,draw,fill=red!3,text width=3.5cm,
			text centered,font=\sffamily,anchor=north, minimum height=2cm},
		leaf5/.style={rectangle,thick,draw,fill=red!3,text width=3.5cm,
			text centered,font=\sffamily,anchor=north, minimum height=3.7cm},
		leaf6/.style={rectangle,thick,draw,fill=red!3,text width=3.5cm,
			text centered,font=\sffamily,anchor=north, minimum height=2.8cm},
		leaf7/.style={rectangle,thick,draw,fill=red!3,text width=3.5cm,
			text centered,font=\sffamily,anchor=north, minimum height=1.1cm},
		edge from parent/.style={very thick,draw=black!70},
		edge from parent path={(\tikzparentnode.south) -- ++(0,-1.05cm)
			-| (\tikzchildnode.north)},
		level 1/.style={sibling distance=14cm,level distance=1.2cm,
			growth parent anchor=south,nodes=root},
		level 2/.style={sibling distance=4cm,level distance=1.2cm},
		level 3/.style={sibling distance=9cm,level distance=1.2cm},
		level 4/.style={sibling distance=5cm,level distance=1.2cm}
		]
		\node (g1) [root] {\textbf{Defense mechanisms for RPL}}
		child {node[l1] (g3) {\textbf{Secure Protocol}}
			child {node[l2] (g4) {\textbf{Cryptography based}}
				child {node[leaf1] (t1) {Dvir \textit{et al.} \cite{Dvir2011}\\Landsmann \textit{et al.} \cite{landsmann2013topology}\\Glissa \textit{et al.} \cite{Glissa}}}
			}
			child {node[l2] (g5) {\textbf{Trust\\ based}}
				child {node[leaf2] (t2) {Seeber \textit{et al.} \cite{seeber2013towards}\\Iuchi \textit{et al.} \cite{iuchi2015secure}\\Airehrour \textit{et al.} \cite{Airehrour2017AJTDE}\\Airehrour \textit{et al.} \cite{Airehrour2017}\\Djedjig \textit{et al.} \cite{Djedjig2017}\\Airehrour \textit{et 
						al.} \cite{Airehrour2018}
					\\Nygaard \textit{et 
						al.} \cite{nygaard2017intrusion}}}
			}
			child {node[l2] (g5) {\textbf{Threshold based}}
				child {node[leaf3] (t2) {Sehgal \textit{et al.} \cite{Sehgal2014}\\Mayzaud \textit{et al.} \cite{Mayzaud2015}
				\\Ghaleb \textit{et al.} \cite{AddressingDAO}}}
			}
		}
		child {node[l1] (g3) {\textbf{Intrusion Detection System}}
			child {node[l2] (g4) {\textbf{Signature based}}
				child {node[leaf4] (t1) {Napiah \textit{et al.} \cite{Napiah2018}\\Kasinathan \textit{et al.} \cite{kasinathan2013ids}
				\\Ioulianou \textit{et al.} \cite{ioulianou2018signature}
				\\Kfoury \textit{et al.} \cite{kfoury2019self}
			}}
			}
			child {node[l2] (g5) {\textbf{Anomaly based}}
				child {node[leaf5] (t2) {Raza \textit{et al.} \cite{RAZA20132661}\\Pongle \textit{et al.} \cite{pongle2015real}\\Mayzaud {et al.} \cite{Mayzaud2016}\\Mayzaud \textit{et al.} \cite{Mayzaud2016Version}\\Shreenivas \textit{et al.} \cite{Shreenivas2017}\\Mayzaud \textit{et al.} \cite{Mayzaud2017}\\Gara \textit{et al.} \cite{gara2017intrusion}\\Gara \textit{et al.} \cite{GaraInderscience}}}
			}
			child {node[l2] (g5) {\textbf{Specification based}}
				child {node[leaf6] (t3) {Le \textit{et al.} \cite{Le2011}\\Zhang \textit{et al.} \cite{zhang2015intrusion}\\Surender \textit{et al.} \cite{Surendar2016}\\Le \textit{\textit{et al.}} \cite{Le2016}\\Shafique \textit{et al.} \cite{shafique2018detection}\\Lai \textit{et al.} \cite{lai2016detection}}}
			}
			child {node[l2] (g6) {\textbf{Hybrid}}
			child {node[leaf7] (t3) {Amin \textit{et al.} \cite{RIDES}
			\\Bostani \textit{et al.} \cite{Bostani2017}}}
		}
		};
		\end{tikzpicture}
	}
	\caption{Taxonomy of defense mechanisms for RPL protocol}
	\label{TOD}
\end{figure*}

\section{Taxonomy of RPL attack defense mechanisms}\label{Taxonomy of RPL attack defense mechanisms}
\textcolor{black}{In this section, different solutions proposed for the detection and mitigation of RPL attacks are discussed.} The solutions present in the literature are divided into two categories: Secure Protocol and Intrusion Detection System (IDS). Secure Protocol based solutions refer to defense mechanisms that are incorporated in the RPL protocol itself, thus, making it secure against various attacks. 
These mechanisms are further categorized into Cryptography, Trust, and Threshold based solutions. \textcolor{black}{Cryptography mechanisms make the use of traditional cryptography methods to provide security and defense against various attacks, whereas trust based mechanisms involve computation of trustworthiness of nodes for facilitating routing decisions.} Threshold based defense solutions exploit the inbuilt feature of RPL and provide an enhancement in order to decide the way trickle timer is reset.  These mechanisms are embedded into RPL protocol, making it more robust in terms of defensive behavior while maintaining desirable network performance. Traditional IDS solutions cannot be directly applied to IoT \cite{Zarpelao2017}. \textcolor{black}{It is because of resource constrained nodes used in the network, different network topologies, and IP based connectivity, which makes traditional IDS solutions infeasible.} This demands for lightweight IDS solutions in terms of computational, communication, memory and energy overhead. In particular to RPL protocol, IDS refers to the second line of defense, which is responsible for the detection of anomalies in RPL operation. These defense solutions can be further classified into Signature, Anomaly, Specification, and Hybrid.

In this section, a brief review of security solutions available in the literature for detecting various attacks in IoT (i.e., typically DoS and RPL based attacks) is presented. Fig. \ref{TOD} shows the taxonomy of various defense solutions, in particular to the RPL protocol.

\subsection{Secure Protocol Based Defense Mechanisms}
This section presents various secure protocol based defense solutions for defending the RPL protocol against routing attacks.      

\subsubsection{Cryptography Based Solutions}

\subsubsection*{Version Number and Rank Authentication (VeRA)}
In \cite{Dvir2011}, a security scheme called VeRA is proposed. \textcolor{black}{The proposed scheme provides defense solutions against attacks related to illegitimate version number and rank change. The key idea is to use hash chains for authenticating those nodes whose rank or version number is changed. VeRa incorporates an authentication mechanism based on hash operations having small time complexity.} The main drawback of VeRA is that it can be bypassed using rank forgery and replay attacks.

\subsubsection*{Enhanced VeRA and Trust Anchor Interconnection Loop (TRAIL)}
\textcolor{black}{To counter Decreased rank attack, Landsmann \textit{et al.} \cite{landsmann2013topology} proposed a novel security mechanism that uses a nested encryption chain to prevent an attacker from multicasting altered hash chains and maintains rank integrity.} The encryption chain links both version number hash chain with rank hash chain. The proposed security mechanism does not provide defense against rank-replay attack. Perrey \textit{et al.} \cite{PerreyLUSW13} proposed an extension to \cite{landsmann2013topology} for detecting and preventing topological inconsistencies. A generic security scheme called Trust Anchor Interconnection Loop (TRAIL) is proposed to facilitate topology authentication in RPL. \textcolor{black}{In TRAIL, each node can validate its upward routing path towards the root and can detect any rank spoofing without relying on encryption chains.} TRAIL can search and remove illegitimate nodes from the network topology. \textcolor{black}{Both VeRA and TRAIL maintain the node's states which incurs memory overhead on resource constrained nodes.}

\subsubsection*{Secure-RPL (SRPL)}
Glissa \textit{et al.} \cite{Glissa} proposed a secure version of the RPL known as SRPL. The main aim of SRPL is to stop compromised nodes from illegitimately manipulating control message information, which may lead to network disruption, i.e., rank manipulation for gaining a better position in the DODAG. SRPL incorporates a security mechanism to maintain a suitable rank threshold such that any change in the rate of rank change leads to the detection of the attack. The rank threshold is implemented with a hash chain authentication of every node in the network. The main advantage of using the proposed solution is that it does not put any limit on node movement from one DODAG to another. When a node moves from one DODAG to another or changing rank, it needs to be validated using secured hashed values at first. SRPL mainly aims to defend Sinkhole, Blackhole, Selective forwarding, and Rank attacks. SRPL involves three phases, namely the initiation phase, the verification phase, and the rank update phase. In the initiation phase, all the nodes in the network compute their rank, threshold values, and respective hashed values. In the verification phase, parents of a respective child node, other nodes check, or verify the hashed rank and thresholds. The rank update is triggered when any node wants to change its rank, and this change is verified against old information and acceptable rank change. The major limitation of SRPL is that it uses computationally expensive operations that consume a lot of node's resources.

\textbf{Summary and Insights:} \textcolor{black}{This section discussed the various cryptography based defense solutions for securing RPL protocol.} It has been observed that the proposed approaches are not sufficient enough to provide the desired security in $ 6 $LoWPAN networks. The proposed solutions face many challenges that need to be addressed. For example, the solution proposed in \cite{Dvir2011} is vulnerable to rank forgery and replay attacks. Similarly, \cite{landsmann2013topology, PerreyLUSW13, Taylor, Glissa} introduce resource overhead (memory, processing), which inhibits their usage in real $ 6 $LoWPAN networks. The approach proposed in \cite{khan2013wormhole} introduces significant communication overhead. In order to leverage the use of cryptography based solutions, further investigation into IoT constraints is needed. Lightweight cryptography solutions can also be explored for developing IoT based security solutions.    

\subsubsection{Trust Based Solutions}

\subsubsection*{Trusted Computing Architecture (TCA)}

\textcolor{black}{In \cite{seeber2013towards}, a  TCA is proposed for establishing trust and facilitating secure key exchange among nodes using a trusted platform module (TPM).} \textcolor{black}{Authors have focused on making the use of low-cost TPM module to incorporate security in resource constrained nodes.} The proposed architecture is capable of defending against node tampering, DoS, and routing attacks targeting availability and integrity. TPM plays a significant role in the proposed architecture as it is responsible for providing keys among authenticated nodes for establishing secure communication. TPM acts as a single point of failure, and if it is tampered or fails, it leads to network performance degradation and security breaches. \textcolor{black}{No extensive evaluation and simulation results have been discussed for validating the effectiveness of TCA.}

\subsubsection*{Secure Parent Selection}
Iuchi \textit{et al.} \cite{iuchi2015secure} proposed a Trust based threshold mechanism for securely selecting a legitimate node as a preferred parent and defending against Rank attacks. In the proposed mechanism, every node in the network selects its preferred parent by assuming the fact that illegitimate node claims a much lower rank than legitimate nodes. All the nodes in the network are capable of finding the illegitimate ranked node by computing the maximum and average rank of its neighbor nodes. A legitimate node then selects its parent node by excluding the node that shows a deficient rank and avoids forwarding packets to illegitimate nodes. \textcolor{black}{The proposed mechanism shows two major limitations. First, it may sometime lead to the creation of unoptimized routes because the legitimate nodes are not selected as a parent in some cases.} Second, the proposed approach is vulnerable to Sybil and Blackhole attacks.

\subsubsection*{Lightweight Trust-Aware RPL}

Ariehrour \textit{et al.} \cite{Airehrour2017AJTDE} proposed a Trust-Aware RPL routing protocol to detect Blackhole and Selective forwarding attacks. The primary idea behind the proposed work is that the packet drop rate of malicious nodes is higher compared to non-malicious nodes when an attacker is performing a Blackhole or Selective forwarding attack. This behavior of nodes is used to evaluate their trustworthiness. The proposed RPL enhancement uses trust values to evaluate the trustworthiness of nodes for facilitating optimal routing decisions. In Trust-Aware RPL initially, all the nodes perform normal path selection operations, i.e., computing route quality over different neighbors based on MRHOF. Trust-Aware RPL shows better performance as compared to MRHOF-RPL in terms of attacks detected, the frequency of node rank changes, throughput, and packet loss. Several drawbacks of the proposed protocol are: (1) promiscuous mode operation increases energy consumption; (2) a legitimate node may begin to drop packets due to unintentional errors that would resemble it as a blackhole attacker.         

\subsubsection*{\textit{\textit{SecTrust}}-RPL}
In \cite{Airehrour2018}, a time based trust aware variant of RPL protocol known as \textit{SecTrust}-RPL is proposed. The proposed RPL variant incorporates a secure trust system that promotes secure communication, detection, and isolation of malicious nodes performing rank and Sybil attacks. \textcolor{black}{The proposed trust mechanism defines a way so that each node in the network computes the trustworthiness of its neighbors by using direct and recommended trust values.} \textit{SecTrust}-RPL incorporates five modules. Trust calculation module is responsible for calculating the trust values of nodes. Trust monitoring module is responsible for updating the trust values of nodes in a periodic and reactive manner. The trust rating process is responsible for sorting trust values in descending order. Detection and isolation of attacks process responsible for selecting high-quality routes and detecting malicious and misbehaving nodes using trust values for ensuring the CIA as well as authenticity. \textcolor{black}{Trust backup and recuperation process take care of the selfish nodes, i.e., nodes which aim to preserve their resources and considered malicious. \textit{SecTrust}-RPL is compared with MRHOF-RPL, and it is shown that the proposed mechanism performs better in terms of attack detected, packet loss, throughput, and frequency of node rank changes.} \textit{SecTrust}-RPL requires nodes to operate in a promiscuous mode, which consequently leads to heavy energy consumption and decreased network lifetime. 

\subsubsection*{Metric based RPL Trustworthiness Scheme (MRTS)}
A trust based security scheme named as MRTS is proposed in \cite{Djedjig2017} for setting up secure routing paths. It works during RPL topology construction and management by incorporating trustworthiness among nodes. In order to perform a secure operation, MRTS defines a new trust based metric named as Extended RPL Node Trustworthiness (ERNT)  and a new trust based objective function named as Trust Objective Function (TOF). ERNT is incorporated in DIO messages and exchanged with neighbor nodes. It is responsible for evaluating the trust value of each node and then quantifies the cost of routing paths. \textcolor{black}{TOF defines a way for nodes to use ERNT and constraints for selecting the preferred parent, and compute their own rank. TOF finds the best routing paths while avoiding the paths with less trustable nodes.} \textcolor{black}{MRTS requires TPM for securing RPL control messages and performs all the security-related computations.} MRTS shows better performance as compared to traditional RPL. However, the main limitations of MRTS are that it uses TPM, which introduces a single point of failure in the network and adds extra hardware cost to the network.

\subsubsection*{Trust based Security System (TIDS)}
Nygaard \textit{et al.} \cite{nygaard2017intrusion} proposed a novel trust-based security system named as TIDS for detecting Sinkhole and Selective forwarding attacks. \textcolor{black}{TIDS enables the normal node to monitor and evaluate its neighbors in order to find anomalies in the normal RPL operation.} The observed data by the node is sent to root (gateway) using Trust Information (TRU) messages for further analysis. The main functionality of TIDS is based on computing trust values using subjective logic. These values are categorized into belief, disbelief, and uncertainty. The trust values are used to analyze the monitored data received from nodes. TIDS is able to detect all the attackers in the network on the cost of heavy energy consumption by the root node and false positives. TIDS requires approximately $ 5 $Kb-$ 6.4 $Kb of ROM and $ 0.7 $Kb-$ 1 $Kb of RAM. The main advantage of the TIDS scheme is that the normal nodes with IDS implemented on it consume very little energy while showing approximately $ 100\% $ detection rate.             

\textbf{Summary and Insights:} It is observed that some solutions present in the literature face a single point of failure issue \cite{seeber2013towards, Djedjig2017}. \textcolor{black}{The solution proposed in \cite{iuchi2015secure} is vulnerable to frequent attacks like Sinkhole and Blackhole.} Several works \cite{Airehrour2017AJTDE, Airehrour2018, nygaard2017intrusion} require nodes to operate in a promiscuous mode which leads to substantial energy consumption. \textcolor{black}{The energy consumption parameter must be considered as the most critical metric while designing any security algorithm for RPL.} Also, the assumption of static networks also adds to one of the essential limitations of work proposed in the literature. These challenges must be addressed before the utilization of proposed solutions in the real network.

\subsubsection{Threshold Based Solutions}

\subsubsection*{Adaptive Threshold (AT)}
In \cite{Sehgal2014}, a mechanism named as Adaptive Threshold (AT) is presented for countering DODAG inconsistency attacks in RPL. The default mechanism (Fixed Threshold) embedded in RPL has a threshold value of $ 20 $. After receiving a packet with `O' and `R' flags set, a node drops the packet and resets the trickle timer. \textcolor{black}{When this number reaches up to a threshold limit of $ 20 $, all such incoming packets are dropped, but the trickle timer is not reset in order to limit the effect of an attack.} \textcolor{black}{This counter is reset after every hour, and in this way, RPL counters the DODAG inconsistency attack.} However, a smart attacker can send $ 20 $ malformed packets every hour and affect the network performance gradually. \textcolor{black}{An attacker can also use different attack patterns to degrade network's performance without getting detected.} AT mechanism considers the current network state to update the threshold based on the rate of receiving packets. \textcolor{black}{The value of threshold decreases when an attacker sends malformed packets very quickly, and increases when an attacker stops sending malformed packets. AT requires prior calculation of optimal configuration parameter values in the arbitrary way (i.e., $ \alpha $, $ \beta $ and $ \gamma$) and does not consider the node mobility.}

\subsubsection*{Dynamic Threshold (DT)}
Mayzaud \textit{et al.} \cite{Mayzaud2015} proposed an improvement to their previous DODAG inconsistency mitigation mechanism \cite{Sehgal2014}. The proposed defense mechanism is known as Dynamic Threshold (DT). \textcolor{black}{It is a fully dynamic threshold mechanism that takes into account the dynamic characteristics of the network to set a threshold for mitigating the DODAG inconsistency attack efficiently. DT does not require any prior calculation of optimal value of configuration parameters like that of AT mechanism because all required information is gathered from network characteristics itself.} It takes into account the convergence time of the network, i.e., the time required by the RPL network to converge. \textcolor{black}{DT approach avoids unnecessary resetting of trickle timer, which consequently suppresses extra DIO transmissions.} DT mechanism outperforms AT mechanism in terms of energy consumption, \textit{PDR}, and end-to-end delay. In addition, the DT mechanism is capable of mitigating the Forced blackhole problem efficiently.

\subsubsection*{SecRPL}
\textcolor{black}{Ghaleb \textit{et al.} \cite{AddressingDAO} proposed SecRPL to address the DAO falsification attack.} The proposed defense mechanism is based on putting a threshold on the number of DAO packets forwarded to each destination. In SecRPL, each parent node maintains a table that contains a counter, specific to every child node in its sub-DODAG. \textcolor{black}{Once the number of DAOs from any child node exceeds the fixed threshold, then that child is marked as malicious.} The parent node drops any further DAO containing the prefix of that malicious child. In order to avoid the situation where any child is permanently blocked, the counter table is reset on every DIO multicast. SecRPL shows significantly good results in terms of the number of DAOs forwarded, control packet overhead, average power consumption, upward, and downward latency. \textcolor{black}{SecRPL requires the selection of optimal threshold limit for efficient operation, which incurs overhead to the security scheme.}     

\textbf{Summary and Insights}: \textcolor{black}{As far as the literature is concerned, only a few works \cite{Sehgal2014, Mayzaud2015, AddressingDAO} focus on using threshold based solutions are available.} Moreover, the proposed solutions address only DODAG inconsistency, Forced blackhole, and DAO falsification attacks, which leaves a big gap to be filled in this field. In addition, the proposed solutions do not consider node mobility, which may hinder the overall system's performance. The key to threshold based solutions lies in the optimal selection of thresholds, i.e., parameters while considering the network environment. This assumption makes such solutions challenging to be developed for other routing attacks. The standard RPL parameters can be used in the optimal selection of thresholds for the development of lightweight threshold based defense solutions \cite{TENCON2019,vermaETT}.

\subsection{Intrusion Detection System (IDS)}
This section discusses various IDS based defense solutions for detecting routing attacks against RPL protocol. IDS based RPL defense mechanisms are summarized in Table \ref{SOIDS}.

\subsubsection{Signature Based IDS}

\subsubsection*{Intrusion Detection System for 6LoWPAN networks}
Kasinathan \textit{et al.} \cite{kasinathan2013ids} proposed an IDS to detect DoS attacks in $ 6 $LoWPAN network . An open-source IDS Suricata is used for pattern matching and attack detection. An IDS probe node is used to sniff all the packet transmissions in the network, and transfer information to Suricata IDS (Open source IDS) for further analysis and attack detection. To prevent communication overhead, the IDS probe node is connected directly to Suricata IDS using a wired link. In addition, a Frequency Agility Manager (FAM) is incorporated to make the network aware of channel occupancy in real-time and operates when the interference level exceeds the set threshold. In this situation, FAM changes the operating channel to the best available one, thus, providing uninterpreted network operations. No simulation study is done in support of IDS performance and its usability.  

\subsubsection*{Compression Header Analyzer Intrusion Detection System (CHA-IDS)}
Napiah \textit{et al.} \cite{Napiah2018} proposed a centralized IDS named CHA-IDS for detecting HELLO flood, Sinkhole and Wormhole
attacks. It uses compression header data to extract certain important network features that are used for detecting individual and combined attacks. \textcolor{black}{The proposed IDS uses the best first and greedy stepwise strategy with correlation-based feature selection to determine the significant features.} \textcolor{black}{Then the selected features are evaluated using six Machine Learning (ML) algorithms (Decision Trees (J48), Logistic Regression (LR), Multi-layer Perceptron (MLP), Naive Bayes (NB), Random Forest (RF), and Support Vector Machine (SVM)) which are used to perform classification of normal and benign traffic.} CHA-IDS outperforms SVELTE and the IDS proposed in \cite{pongle2015real}. The main limitations of CHA-IDS include high memory and energy consumption. Moreover, it is incapable of identifying the attacker. 

\subsubsection*{Signature-based Intrusion Detection System}
A framework for a signature-based IDS to detect DIS and Version number attack is proposed in \cite{ioulianou2018signature}. The proposed IDS requires detection and monitoring modules to be placed on nodes itself, as in the case of hybrid detection schemes. \textcolor{black}{However, the authors consider two types of additional nodes in the proposed scheme.} The first type of nodes are IDS routers, which carry detection and firewall modules. The second type of nodes are sensors or IDS detectors which are responsible for monitoring and sending malicious traffic information to the router nodes. IDS router checks all the passing traffic to decide whether the packet source is malicious or not. The job of the IDS detector is to monitor sensor traffic and calculate the metric of interest \textcolor{black}{, i.e.,  Received Signal Strength Indicator (RSSI)}, packet drop rate, and packet sending rate. The final decision of classifying a node as malicious or not is taken by detection module running on $ 6 $BR, based on the data received from each node. The proposed framework is not validated, which is its major limitation. 

\begin{landscape}
	\centering
	\begin{table}
		\small
		\caption{Summary of Secure Protocol based defense mechanisms}
		\label{SecureProtocol}
	\color{black}\begin{tabular}[l]{|p{2.2cm}|p{2.5cm}|p{4cm}|p{5cm}|p{0.3cm}|p{0.3cm}|p{1.6cm}|p{4cm}|}
			
			\hline
			\multicolumn{1}{|c|}{\textbf{Reference}} & \multicolumn{1}{c|}{\textbf{Defense Mechanism}} & \multicolumn{1}{c|}{\textbf{Relevant Attack}} & \multicolumn{1}{c|}{\textbf{Limitations}} & \multicolumn{1}{c|}{\textbf{Mobility}} & \multicolumn{1}{c|}{\textbf{Validation}} &
			\multicolumn{1}{c|}{\textbf{Tools/Motes}} &  \multicolumn{1}{c|}{\textbf{Performance metrics}} \\ \hline
			%
			Dvir \textit{et al.} \cite{Dvir2011} & VeRA & Version number and Decreased rank & Vulnerable to Rank-replay attack, Hash chain forgery attack, adds memory and computational overhead. & No & - & - & - \\ \hline
			Landsmann \textit{et al.} \cite{landsmann2013topology} & Enhanced VeRA & Version number and Decreased rank & Vulnerable to Rank-replay attack, adds memory overhead, child node might select attacker as a parent. & No & - & - & - \\ \hline
			Perrey \textit{et al.} \cite{PerreyLUSW13} & TRAIL & Version number, Decreased rank, Rank-replay & Adds memory overhead. & No & Testbed & DES Mesh/RIOT OS & Routing convergence time, Average message size \\ \hline

			Seeber \textit{et al.} \cite{seeber2013towards} & Trusted Computing Architecture & RPL routing attacks targeting availability and integrity, node tampering & TPM is a single point of failure,  adds computational overhead due to  cryptography processing. & No & - &-  & - \\ \hline
			Sehgal \textit{et al.} \cite{Sehgal2014} & Adaptive Threshold & DODAG inconsistency & Requires prior calculation of configuration parameters (optimal values). & No & Simulation & Contiki OS/Cooja & \textit{PDR}, Energy consumption and Control packet overhead \\ \hline
			Mayzaud \textit{et al.} \cite{Mayzaud2015} & Dynamic Threshold & DODAG inconsistency & Increases energy consumption. & No & Simulation & & Control packet overhead, \textit{PDR}, Energy consumption \\ \hline
			
			Ghaleb \textit{et al.} \cite{AddressingDAO} & SecRPL & DAO falsification & Increases Average power consumption, Control packet Overhead and Latency. Decreases \textit{PDR} and degrades network reliability. & No & Simulation & Contiki OS/Cooja & Control Packet Overhead, \textit{PDR}, Energy consumption, DAO forwarding overhead, Upward and downward Latency \\ \hline

			Iuchi \textit{et al.} \cite{iuchi2015secure} & Secure Parent Selection & Rank & Susceptible to Sybil and Blackhole attacks,  may result in longer paths (unoptimized). & No & Simulation & Contiki OS/Cooja & Total number of child nodes attached to attacker nodes. \\ \hline

			Glissa \textit{et al.} \cite{Glissa} & Secure-RPL (SRPL) & Rank, Sinkhole and Selective forwarding attacks & Computationally expensive. & No & Simulation & Contiki OS/Cooja & Average power consumption, Control message overhead, and Packet reception rate \\ \hline
			Djedjig \textit{et al.} \cite{Djedjig2017} & Metric-based RPL Trustworthiness Scheme (MRTS). & Insider attacks & Adds computation and communicate overhead and increases energy consumption. & No & - & - & - \\ \hline
			Ariehrour \textit{et al.} \cite{Airehrour2017AJTDE} & Trust-Aware RPL & Blackhole and Selective forwarding & Nodes need to operate in promiscuous mode to overhear neighbor transmissions which adds energy overhead. & No & Simulation & Contiki OS/Cooja & Detection rate, Throughput, Packet loss and Frequency of node rank changes. \\ \hline
			Ariehrour \textit{et al.} \cite{Airehrour2017} & Trust-Aware RPL for detecting Blackhole & Blackhole & Nodes need to operate in promiscuous mode which adds energy overhead. & No & Testbed & Contiki/ XM1000 motes & Detection rate, Throughput, Packet loss and Frequency of node rank changes. \\ \hline
			Ariehrour \textit{et al.} \cite{Airehrour2018} & SecTrust-RPL & Rank and Sybil & Considers static network topology, nodes need to operate in promiscuous mode which increases energy consumption. & No & Simulation and Testbed & Contiki/ XM1000 motes & Detection rate, Throughput, Packet loss and Frequency of node rank changes. \\ \hline
			Nygaard \textit{et al.} \cite{nygaard2017intrusion} & TIDS & Sinkhole and Selective forwarding & Considers static network topology, requires $ 6 $BR (root) to remain constantly ON which consequently increases energy consumption, high \textit{FPR}. & No & Simulation & Contiki OS/Cooja & \textit{Detection rate, FN, FP}, Energy consumption \\ \hline
		\end{tabular}
	\end{table}
\end{landscape}

\subsubsection*{Self Organizing Map Intrusion Detection System (SOMIDS)}

Kfoury \textit{et al.} \cite{kfoury2019self} proposed SOMIDS for detecting Sinkhole, Version number, and HELLO flooding attacks. SOMIDS uses Self Organizing Maps (SOM) for clustering attacks and normal traffic. \textcolor{black}{SOMIDS uses a Pcap file from a cooja simulator for extracting data and performing clustering of traffic classes.} SOMIDS consists of three major components. The first component is an aggregator module that is responsible for aggregating the data (ICMPv$ 6 $ code, IPv6 destination, IPv$ 6 $ source, ICMPv$ 6 $ DIO version, ICMPv$ 6 $ DIO rank, Timestamp) contained in captured PCAP file. Traffic data is aggregated into six variables, i.e., number of DIS, DIO, DAO messages, the ratio of version number changes, the ratio of rank changes, and average mote power. The second component is normalizer, which performs the task of normalizing the aggregated data. \textcolor{black}{Third component is a trainer module which is responsible for training SOM. The result of the IDS is a matrix that is converted into a $ 2 $D image for better visualization of clusters.} SOMIDS is not evaluated in terms of the implementation overhead and does not consider node mobility.

\textbf{Summary and Insights:} It is analyzed that some of the proposed approaches \cite{kasinathan2013denial, kasinathan2013ids} rely on the outdated signatures (traffic patterns) for classifier training which makes these approach ineffective for securing RPL networks. \textcolor{black}{The solutions proposed in \cite{Napiah2018,kfoury2019self, shukla2017ml} used signatures collected from the simulated attacks.} These approaches show promising results in terms of prominent metrics. However, signatures collected from the real network can be more effective in classifier training. The development of RPL based real traffic dataset containing traces of common routing attacks needs to be done \cite{verma2019evaluation,abhishek_verma_IoTSIU}. The signature based IDS proposed in \cite{Napiah2018} can be improved in terms of energy consumption. 

\subsubsection{Anomaly Based IDS}

\subsubsection*{SVELTE}
Raza \textit{et al.} \cite{RAZA20132661} proposed a real-time IDS named SVELTE for $ 6 $LoWPAN. The proposed IDS consists of anomaly based detection engine which uses RPL specifications for detecting spoofed information, Sinkhole, and Selective forwarding attacks. \textcolor{black}{It consists of three centralized modules that are placed on 6BR: Mapper, Analyzer and Detector, and a Mini-firewall.} Every child node sends RPL information to $ 6 $BR for illegitimate traffic filtering. Intrusion detection in SVELTE involves network graph inconsistency detection, node availability detection, and routing graph validation. SVELTE imposes very less memory, computational, and energy overhead on the resource constrained nodes. Moreover, it shows a good performance in terms of \textit{PDR} and control packet overhead. \textcolor{black}{The limitations of SVELTE include strategic placement of IDS modules, timing inconsistency in rank measurements, which consequently leads to inaccurate topology creation at 6BR, and high false positive rate (\textit{FPR}). In addition, SVELTE does not provide defense against coordinated attacks. }

\subsubsection*{Real Time Intrusion and Wormhole Detection}
A novel IDS for the detection of Wormhole attack in IoT is proposed in \cite{pongle2015real}. It detects the packet relay and encapsulation types of Wormhole attack. The proposed IDS uses the node's location and neighbor information to identify the attack and received signal strength indicator (RSSI) to identify attacker nodes. A hybrid deployment strategy on a static network is considered for placing IDS modules, where a centralized module is placed on 6BR, and distributed modules are placed on resource constrained nodes. Distributed modules are responsible for sending and monitoring RSSI values, sending neighbor information to 6BR, and packet forwarding. Centralized modules collect RSSI values, compute the distance from the node's RSSI value, and perform validation of neighbors from collected information and detect attack with its location.  The main drawback of the proposed IDS is that it puts much communication and computational burden on resource constrained nodes.

\subsubsection*{Distributed Monitoring Architecture}
Mayzaud \textit{et al.} \cite{Mayzaud2016} proposed a distributed monitoring architecture for detecting DODAG inconsistency attacks. \textcolor{black}{The proposed architecture makes the use of RPL multi-instance feature and dedicated monitoring nodes for facilitating energy efficient network events observation (passively).} Two types of nodes are considered in the network, i.e., regular (monitored) and monitoring nodes. The multi-instance feature of RPL is used for creating regular (the network of regular nodes) and monitoring network (the network of monitoring nodes). The monitoring nodes contain local anomaly detection (algorithm) modules that analyze the collected data and detect possible attacks in a distributed manner. The main limitations of the proposed architecture are: it assumes a single attacker case and fails in case of multiple attackers which are operating in a collaborative manner, monitoring nodes need to operate in promiscuous modes for anomaly detection, depends on the coverage of regular nodes by monitoring nodes (strategic placement), relies on high order devices for monitoring which adds cost overhead, architecture relies on local detection. 

\subsubsection*{Extension to Distributed Monitoring Architecture}

Mayzaud \textit{et al.} extended their previous proposed approach \cite{Mayzaud2016} in \cite{Mayzaud2016Version} to detect Version number attacks. Authors considered the fact that an incremented version number is propagated in the entire graph, and a monitoring node cannot decide by itself if this is the result of an attack or not, and they must share monitoring information to identify the malicious node more efficiently. Thus, they extended the distributed monitoring architecture such that monitoring nodes can collaborate together using a multi-instance network and facilitate global detection. \textcolor{black}{Only one attacker case is assumed, and mobility is not considered in this defense architecture.} An extension to \cite{Mayzaud2016Version} is presented in \cite{Mayzaud2017}. In this work, detection and localization algorithms are presented. The \textit{``LOCAL\_ASSESSMENT"} algorithm is deployed on monitoring nodes except the root, which allows monitoring nodes to report to the root the sender of an incremented version number in their neighborhood. The \textit{``DISTRIBUTED\_DETECTION"} algorithm is deployed on the sink to detect the attack and gather all monitoring node information into tables. The \textit{``LOCALIZATION"} algorithm is deployed on the sink node and performs attacker identification by analyzing the collected information. This framework inherits the limitations of Mayzaud \textit{et al.} \cite{Mayzaud2016}.         

\subsubsection*{Extended SVELTE based on ETX metric}
An extension to SVELTE is proposed in \cite{Shreenivas2017}. In addition to SVELTE IDS modules, an extra intrusion detection module which uses the ETX metric is incorporated for the detection of ETX manipulation attacks in ETX metric based RPL networks. The authors have also proposed an intrusion detection method which uses geographical parameters (node's location and transmission limits) for handling a case when both rank and ETX based detection methods fail. The main idea behind ETX based intrusion detection method is that ETX value of the parent node must be lower than that of its children node, and if any node's ETX value is found to be inappropriate or unusual, then the node reported as malicious. The main advantage associated with the proposed solution is that the ETX based IDS can defend against ETX and rank based attacks. In contrast, the geographical parameter based method can locate the nodes and test their authenticity. The proposed IDS solutions consume less power when nodes operate in duty cycling mode and require only $ 5,570 $ and $ 6 $ Bytes of RAM and ROM, respectively. A high true positive rate (\textit{TPR}) is achieved when both the proposed solutions are combined together. The proposed solution does not consider node mobility in the network.

\subsubsection*{Hybrid IDS based on the Sequential Probability Ratio Test with an Adaptive Threshold}
A hybrid IDS that combines the Sequential Probability Ratio Test (SPRT) with an Adaptive Threshold to detect Selective forwarding attack is proposed in \cite{gara2017intrusion}. It uses two types of modules, a centralized module deployed on the gateway node and a distributed module deployed on resource constrained nodes. The proposed IDS involves three steps, i.e., data gathering, data analysis, decision, and elimination of compromised node. The data gathering step involves each routing node to collect the neighbor's information, storing it in the form of a table, and then send it to the centralized node using HELLO messages. The data analysis step involves the computation of the number of dropped packets and the probability of dropped packet for each node using data gathered from HELLO messages. The decision step is responsible for detecting malicious nodes and minimizing \textit{FAR} by utilizing SPRT. The elimination of the compromised node step involves informing legitimate nodes about the compromised nodes by initiating a global repair and sending the compromised node's identifier in fresh DIO messages to all other legitimate nodes in the network.       
The proposed IDS achieves $ 100\% $ detection rate. However, the communication overhead of the network increases with the increase in node mobility. 

\textbf{Summary and Insights:} Many of the anomaly based IDS solutions present in the literature show acceptable performance, which favors their utility in IoT applications. However, it is observed that the proposed solutions achieve high performance (accuracy, \textit{TRP, FPR}, etc.) while imposing an additional cost to the nodes in terms of communication, computation, memory, and energy consumption. \textcolor{black}{The solutions proposed in \cite{pongle2015real, Cervantes2015, Mayzaud2016, Mayzaud2016Version, chen2016defense, gara2017intrusion ,GaraInderscience} impose extra network deployment cost which is undesirable for resource constrained networks.} Similarly, the security approach proposed in \cite{RAZA20132661} requires the strategic placement of IDS monitoring modules, which add an implementation complexity to the network. \textcolor{black}{Moreover, it is also observed that the proposed anomaly based IDS are still vulnerable to the coordinated attacks.} These critical challenges must be addressed for the advanced development of anomaly based IDS for IoT.      
\subsubsection{Specification Based IDS}

\subsubsection*{Intrusion detection and response system for Internet of things (InDReS)}
In \cite{Surendar2016}, a distributed IDS named InDReS to detect Sinkhole attack in RPL is proposed. The proposed IDS is based on cluster tree topology, where cluster head acts as a monitoring node that observes packet drop count of its adjacent nodes. \textcolor{black}{The monitoring nodes compute the rank of every adjacent node to it and compare that rank with the threshold value for finding a malicious node. InDReS is implemented on NS-$ 2 $, and performance results are compared with that of INTI. The results show that the proposed IDS performs well compared to INTI in terms of packet drop ratio, \textit{PDR}, control packet overhead, and average energy consumption. The limitations of InDReS include: only homogeneous nodes are considered, the dynamic network is not considered, and it may fail if the leader node itself gets compromised.}

\subsubsection*{Specification-Based IDS for Detecting Topology Attacks}
\textcolor{black}{Le \textit{et al.} in their previous work \cite{Le2011} proposed a specification based IDS architecture which lacks implementation and performance analysis.} In \cite{Le2016}, the authors extended the previous architecture and evaluated it in terms of prominent evaluation metrics. They proposed a specification based IDS consisting of Extended Finite State Machine (EFSM) that is generated from a semi-auto profiling technique. Firstly, EFSM is created from RPL specification using ILP (Integer Linear Programming) technique to define stable states and transitions among them. Secondly, RPL knowledge of the RPL profile of detection algorithms is translated to form more concrete states and transitions, i.e., utilizing trace files generated from RPL normal operation in the Cooja simulator. This specification defines all the legitimate states and transitions which a node must follow while operating in a normal manner. EFSM is implemented as a set of rules on intrusion detection agents for detecting various attacks, including Rank, Local Repair, Neighbor, DIS, and Sinkhole. The proposed IDS is shown to achieve \textit{TPR} of $ 100\% $ with \textit{FPR} up to $ 6.78\% $. The proposed IDS introduces communication overhead, requires a good network trace for the creation of effective specification, and shows less accuracy when it works for a long time. 

\subsubsection*{RPL-Based Wormhole Detection}
Lai \textit{et al.} \cite{lai2016detection} proposed a distributed wormhole detection method which applies the rank information to estimate the relative distance from the root node. The proposed method uses the hop count metric for rank calculation. To detect malicious nodes, the proposed detection method checks for the nodes with unreasonable rank values. It defines \textit{Rank\_Threshold} and \textit{Rank\_Diff} attributes for the detection of illegitimate DIO messages. \textit{Rank\_Threshold} is defined as the difference between the rank values of parent and node itself, whereas \textit{Rank\_Diff} is the difference between the rank values of the source node and node itself. DIO message is considered as abnormal, when \textit{Rank\_Diff}$ > $\textit{Rank\_Threshold} condition is not met. The proposed wormhole detection method shows a $ 100\% $ output in terms of precision, recall, and accuracy. The main advantages of this approach are its easy implementation and no additional requirement for Wormhole attack detection. However, node mobility is node considered, which can severely affect the detection results. In addition, critical parameters like \textit{PDR}, end-to-end delay, and energy consumption are not analyzed.

\subsubsection*{Specification based IDS based on Finite State Machine}
\textcolor{black}{In \cite{Le2011}, a specification based IDS is proposed for detecting rank and local repair attacks.} The proposed IDS uses a finite state machine (FSM) for monitoring the node's state, i.e., normal or malicious. A backbone architecture is used for placing monitoring nodes containing FSM modules. Monitoring nodes sniff neighbor transmissions, including its parent and child nodes. \textcolor{black}{The parameters like node id, the preferred parent with their respective rank, state changes in a specific period are monitored and extracted from sniffed DIO messages in order to analyze the node's behavior.} Monitoring nodes collaborate and share information for detecting attacker nodes. FSM specifies normal and malicious states. \textcolor{black}{FSM state specifies the strict rank rule which nodes must follow, i.e., parent-child relationship, and an acceptable threshold for the number of times a topology can be set up or updated.} Any deviation from the specified rules and threshold consequently changes the node's state from normal to suspicious and detects the possible attacker node.

\subsubsection*{IDS to defense Routing choice intrusion Intrusion}
An IDS to defend against Routing choice intrusion (ETX metric) is proposed in \cite{zhang2015intrusion}. The proposed IDS is based on specification methodology that uses a stand alone architecture with distributed monitoring nodes. Authors consider attack defense only against a single intruder case. The proposed IDS requires monitoring nodes containing FSM with normal and malicious states. \textcolor{black}{Network behaviors are matched with FSM states, and any deviation from normal state leads to attack detection. Routing choice intrusion is detected in the case when any malicious node multicast the DIO with lower ETX value, which consequently leads to a large fluctuation in the number of its child nodes than a set threshold, this node is marked as an attacker node.} The authors consider certain assumptions like secure network initialization, homogeneous nodes, monitoring nodes with more resources, and static environment, which limits the practicality of the proposed IDS.

\subsubsection*{Sink-based Intrusion Detection System (SBIDS)}
\textcolor{black}{In \cite{shafique2018detection}, a centralized specification based IDS known as SBIDS is proposed to address rank attacks in RPL based IoT networks.} SBIDS uses information contained in the DAO message received from child nodes in its sub-DODAG. SBIDS utilizes RPL parameters, including node's current rank (NCR), node's parent rank (NPR), node's previous rank (NPVR), and parent switching threshold (PST) for detecting whether a node is malicious or not. SBIDS achieves $ 100\% $ accuracy in case of a static network. \textcolor{black}{The accuracy decreases in the presence of mobile nodes in the network.} SBIDS adds a communication overhead to RPL protocol as it requires an extra $ 48 $-bit information to be added by the nodes in the DAO packets they send. SBIDS shows better results for a static network as compared to the mobile network. The average power consumption of nodes increases in the case of SBIDS.

\textbf{Summary and Insights:}  The effectiveness of specification based IDS solutions can be observed from their performance. The only key challenge in the development of specification based IDS is the availability of quality traffic trace required for generating adequate specifications \cite{Le2016}. \textcolor{black}{It is observed that several approaches \cite{lai2016detection, shafique2018detection} have not performed power consumption analysis, hence there exists an open research gap to be considered for future research}. Moreover, the integration of mobility support in the proposed solutions is a challenging task and needs further investigation.

\subsubsection{Hybrid IDS} 

\subsubsection*{Robust Intrusion Detection System (RIDES)}
Amin \textit{et al.} \cite{RIDES} proposed a novel IDS named RIDES for detecting DoS attacks in IP based WSN. It is a hybrid of signature and anomaly based IDS. The signature based intrusion detection component uses a distributed pattern matching using bloom filters to match signature codes. \textcolor{black}{To reduce the overhead to long signature codes, a coding scheme is used which converts signatures into short attack identifiers.} The anomaly based intrusion detection component uses Cumulative Sum Control charts (CUSUM) with upper and lower threshold limits to detect anomalies in the network pattern. A distributed approach is used to place the intrusion detection components for decreasing the communication, memory, and computational overhead on nodes. The main limitation of this work is inter-packet delay that leads to delayed intrusion detection by RIDES. In addition to it, energy consumption by the resource constrained nodes is not studied.

\subsubsection*{Hybrid of Anomaly and Specification based on optimum-path forest clustering} 
A novel real-time hybrid IDS framework is proposed in \cite{Bostani2017} to detect Sinkhole, Selective forwarding, and Wormhole attacks. Specification based IDS modules are deployed on router nodes which perform analysis of their child nodes and forward their local results to the gateway node through data packets. The gateway node is equipped with anomaly based IDS module which employs Unsupervised Optimum-Path Forest Clustering (OPFC) algorithm for projecting clusters by using incoming data packets. \textcolor{black}{The simulation results show that the proposed IDS framework achieves the maximum \textit{TPR} of $ 96.02\% $ with $ 2.08\% $ of \textit{FPR}.} The main features of the proposed hybrid IDS include high scalability and attacker identification. There are several drawbacks associated with this hybrid IDS. It does not consider the energy constrained nature of nodes, assumes one-way communication (node to gateway), and considers only a static network. 

\textbf{Summary and Insights:} Similar to signature and anomaly based IDS, hybrid based IDS solutions also face several challenges that need to be addressed. Delayed attack detection makes IDS solutions inefficient when deployed in real networks. The IDS proposed in \cite{RIDES} is affected by the inter-packet delay that causes delayed attack detection. Such issues need to be carefully addressed while designing IDS for IoT applications. Hybrid IDS proposed by Bostani \textit{et al.} \cite{Bostani2017} utilized MapReduce architecture to manage a large amount of data from motes and perform attack detection efficiently. Other such algorithms available in the literature need to be explored for building scalable and effective IDS solutions corresponding to IoT.

Table \ref{tab:PerfComp} presents a comparative study of discussed security solutions (Secure Protocol and IDS) based on different evaluation metrics. The performance is compared based on the maximum improvements achieved in percentages (\%), and maximum or minimum values (val) achieved. 

\onecolumn
\color{black}\begin{landscape}
	\centering
	\small
	\begin{longtable}{|p{1.8cm}|p{2.2cm}|p{1.45cm}|p{1.4cm}|p{2.3cm}|p{5.5cm}|p{1.15cm}|p{1.3cm}|p{1.7cm}|p{2cm}|}
		
		\caption{Summary of Intrusion Detection System based defense mechanisms}
		\label{SOIDS}\\
		\hline
		\textbf{Reference} & \textbf{Defense Mechanism} & \textbf{Type} & \textbf{Placement strategy} & \textbf{Relevant Attack} & \textbf{Limitations} & \textbf{Mobility} & \textbf{Validation} &
		\multicolumn{1}{c|}{\textbf{Tools/Motes}} & \textbf{Performance metrics} \\ \hline
		\endhead
		Amin \textit{et al.}\cite{RIDES} & RIDES & Hybrid & Distributed & DoS & Inter packet delay affects the detection time. & No & Simulation & ns-2 & \textit{TPR, FPR, ROC} \\ \hline
		Le \textit{et al.}\cite{Le2011} & Specification based IDS & Specification & Distributed & Rank, Local repair & No simulation study has been done for the proposed IDS. & No & - & -  & - \\ \hline
		Raza \textit{et al.}\cite{RAZA20132661} & SVELTE & Anomaly & Hybrid & Sinkhole, Selective forwarding, spoofed or altered information & Synchronization issue, requires strategic placement of IDS modules, high FPR, vulnerable to coordinated attacks. & No & Simulation & Contiki OS /Cooja & Energy consumption, \textit{TPR} \\ \hline

		Kasinathan \textit{et al.}\cite{kasinathan2013denial} & DoS detection IDS Architecture & Signature & Centralized & DoS & The centralized nature of the IDS architecture makes it difficult to detect internal attacks and introduces communication overhead over resource constrained nodes. & No & Testbed & PenTest /Contiki OS& \textit{TP} \\ \hline
		Kasinathan \textit{et al.}\cite{kasinathan2013ids} & Intrusion Detection System for 6LoWPAN networks & Signature & Centralized & DoS & The centralized nature of the IDS architecture makes it difficult to detect internal attacks and introduces communication overhead over resource constrained nodes. & No & Testbed & PenTest /Contiki OS& - \\ \hline

		Zhang \textit{et al.}\cite{zhang2015intrusion} & IDS to defense Routing choice Intrusion & Specification & Distributed & Routing choice intrusion & Assumes secure network initialization and homogeneous devices. Monitoring nodes need to operate in promiscuous mode. & No & Simulation & Contiki OS /Cooja & - \\ \hline
		Pongle \textit{et al.}\cite{pongle2015real} & Real Time Intrusion Detection System & Anomaly & Hybrid & Wormhole & Introduces communication and computational overhead. & No & Simulation & Contiki OS /Cooja & \textit{TPR}, Energy consumption, Control packet overhead \\ \hline

		Mayzaud \textit{et al.}\cite{Mayzaud2016} & Distributed Monitoring Architecture & Anomaly & Distributed & DODAG inconsistency & It assumes a single attacker case and fails in case of multiple attackers operating in a collaborative manner. Monitoring nodes need to operate in promiscuous modes for anomaly detection. Depends on the coverage of regular nodes by monitoring nodes (strategic placement). Relies on high order devices for monitoring, which adds cost overhead. Architecture relies on local detection. & No & Simulation  & Contiki OS /Cooja & - \\ \hline
		Mayzaud \textit{et al.}\cite{Mayzaud2016Version} & Distributed Monitoring Architecture & Anomaly & Hybrid & Version number & It considers only a single attacker case, monitoring nodes need to operate in promiscuous modes for anomaly detection, relies on high order devices for monitoring, which adds cost overhead. Do not consider node mobility and depends on the coverage of regular nodes by the monitoring nodes (strategic placement). & No & Simulation & Contiki OS /Cooja & \textit{FPR} \\ \hline
		Mayzaud \textit{et al.}\cite{Mayzaud2017} & Distributed Monitoring Architecture & Anomaly & Hybrid & Version number & It considers only a single attacker case, monitoring nodes need to operate in promiscuous modes for anomaly detection, relies on high order devices for monitoring, which adds cost overhead. Do not consider node mobility and depends on the coverage of regular nodes by the monitoring nodes (strategic placement). & No & Simulation & Contiki OS /Cooja & \textit{FPR} \\ \hline
		Surender \textit{et al.}\cite{Surendar2016} & InDReS & Specification & Distributed & Sinkhole & It considers only homogeneous nodes and do not consider network dynamicity. This approach may fail if leader node itself gets compromised. & No & Simulation & ns-2 & Packet drop ratio, \textit{PDR}, Throughput, Energy consumption, Control packet overhead \\ \hline
		Le \textit{et al.}\cite{Le2016} & Specification based IDS & Specification & Hybrid & Rank, Sinkhole, Local repair, Neighbor, DIS & Introduces communication overhead, requires a good network trace for the creation of effective specification, and shows less accuracy when it works for a long time. & No & Simulation & Contiki OS /Cooja & \textit{TPR, FPR}, Energy consumption \\ \hline
		Lai \textit{et al.}\cite{lai2016detection} & RPL-Based Wormhole Detection & Specification & Distributed & Wormhole & Node mobility is not considered which can severely affect detection results. Critical parameters like  \textit{PDR}, end-to-end delay and energy consumption are not analyzed. & No & Simulation & - & Precision, Recall and Accuracy \\ \hline
		Shreenivas \textit{et al.}\cite{Shreenivas2017} & Extended SVELTE based on ETX metric & Anomaly & Hybrid & ETX manipulation, Rank & Do not consider mobility. Parameters like end-to-end delay, \textit{PDR} are not not analyzed. & No & Simulation & Contiki OS /Cooja & Average power consumption, \textit{TPR} \\ \hline
		Chen \textit{et al.}\cite{chen2016defense} & Intrusion Detection System for Detecting Wormhole and Flooding Attacks & Anomaly & - & Wormhole, Flooding & Overhead of maintaining blacklist in large-scale networks affects overall network performance. Placement strategy for IDS modules is not discussed. & No & Simulation & - & Precision, Recall, Accuracy and Miss rate \\ \hline

		Ahsan \textit{et al.}\cite{ahsan2017wormhole} & ABR-SAR based IDS for Wormhole detection & Anomaly & Hybrid & Wormhole & Increases implementation complexity. Strategic placement of SAN is needed so that every node must be in range of at least one another SAN. & No & Simulation  & Contiki OS /Cooja & Detection rate, Average power consumption \\ \hline
		Gara \textit{et al.}\cite{gara2017intrusion} & Hybrid Intrusion Detection System based on Sequential Probability Ratio Test with an Adaptive Threshold & Anomaly & Hybrid & Selective forwarding & Exchange of HELLO messages increases network overhead. & Yes & Simulation & Contiki OS /Cooja & Detection rate, Control packet overhead \\ \hline
		Gara \textit{et al.}\cite{GaraInderscience} & Hybrid Intrusion Detection System based on Sequential Probability Ratio Test with an Adaptive Threshold & Anomaly & Hybrid & Selective forwarding and Clone ID & Exchange of HELLO messages increases network overhead. & Yes & Simulation &  Contiki OS /Cooja & Detection rate, Control packet overhead \\ \hline
		Napiah \textit{et al.}\cite{Napiah2018} & Compression Header Analyzer Intrusion Detection System (CHA-IDS) & Signature & Centralized & HELLO flooding, Sinkhole and Wormhole & Introduces memory and energy consumption. It cannot identify the attacker. & No & Simulation & Contiki OS/ Cooja/ Weka & \textit{TPR, FPR}, Accuracy, Energy Consumption \\ \hline
		Bostani \textit{et al.}\cite{Bostani2017} & Hybrid of Anomaly and Specification based IDS & Hybrid & Distributed & Sinkhole and Selective forwarding & Assumes one way communication. Energy overhead analysis is not done. & No & Simulation & MATLAB & \textit{TPR, FPR}, Accuracy\\ \hline
		Shafique \textit{et al.}\cite{shafique2018detection} & SBIDS & Specification & Centralized & Rank & Introduces communication overhead and increases Average power consumption. & Yes & Simulation & Contiki OS /Cooja & \textit{TP, FP, FN, FP}, Accuracy, Average power consumption\\ \hline
		Ioulianou \textit{et al.}\cite{ioulianou2018signature} & Framework of Signature-based IDS & Signature & Hybrid & HELLO flooding and Version number & No validation is performed in support of the framework. & No & - & - & -\\ \hline
		Kfoury \textit{et al.}\cite{kfoury2019self} & SOMIDS & Signature & Centralized & HELLO flooding, Sinkhole, and Version number & No evaluation in terms of prominent performance metrics is done. Energy consumption of 6BR is not studied. & No & Simulation & Contiki OS /Cooja /Python & - \\ \hline
		Shukla \textit{et al.}\cite{shukla2017ml} & ML-IDS (KM-IDS, DT-IDS and Hybrid-IDS) & Signature & Centralized & Wormhole & \textit{FP} value is not reported.  Energy consumption and deployment strategy are not discussed.   & No & Simulation & C++ & Detection rate\\ \hline
	\end{longtable}
\end{landscape}

\begin{landscape}
	\centering
	\footnotesize
	\begin{longtable}{|c|c|c|c|c|c|c|c|c|c|c|c|c|c|c|c|c|c|c|c|c|c|c|c|c|c|c|}
		
		\caption{Performance comparison of security solutions in terms of different evaluation metrics}
		
		\label{tab:PerfComp}\\
		\hline
		\multicolumn{1}{|c|}{\textbf{Reference}} & \multicolumn{1}{c|}{\textbf{\rotatebox[origin=c]{90}{Throughput (\%)}}} & 
		\multicolumn{1}{c|}{\textbf{\rotatebox[origin=c]{90}{Routing convergence  time (\%)}}} & \multicolumn{1}{c|}{\textbf{\rotatebox[origin=c]{90}{Extra messages per node (val)}}} & \multicolumn{1}{c|}{\textbf{\rotatebox[origin=c]{90}{Packet delivery ratio (\%)}}} & \multicolumn{1}{c|}{\textbf{\rotatebox[origin=c]{90}{Energy consumption (\%)}}} & \multicolumn{1}{c|}{\textbf{\rotatebox[origin=c]{90}{Control packet overhead (\%)}}} & \multicolumn{1}{c|}{\textbf{\rotatebox[origin=c]{90}{DAO forwarding overhead (\%)}}} & \multicolumn{1}{c|}{\textbf{\rotatebox[origin=c]{90}{Upward latency (\%)}}} & \multicolumn{1}{c|}{\textbf{\rotatebox[origin=c]{90}{Downward latency (\%)}}} &
		\multicolumn{1}{c|}{\textbf{\rotatebox[origin=c]{90}{Average power consumption (\%)}}} & \multicolumn{1}{c|}{\textbf{\rotatebox[origin=c]{90}{Packet reception ratio (\%) }}} &
		\multicolumn{1}{c|}{\textbf{\rotatebox[origin=c]{90}{Packet loss (\%)}}} & \multicolumn{1}{c|}{\textbf{\rotatebox[origin=c]{90}{Node rank changes (\%)}}} &
		\multicolumn{1}{c|}{\textbf{\rotatebox[origin=c]{90}{True negatives (\%)}}} & \multicolumn{1}{c|}{\textbf{\rotatebox[origin=c]{90}{False negatives (\%) }}} & \multicolumn{1}{c|}{\textbf{\rotatebox[origin=c]{90}{False positives (\%)}}} & 
		\multicolumn{1}{c|}{\textbf{\rotatebox[origin=c]{90}{False positive rate (val)}}} & \multicolumn{1}{c|}{\textbf{\rotatebox[origin=c]{90}{Receiver operating characteristic (val)}}} &
		\multicolumn{1}{c|}{\textbf{\rotatebox[origin=c]{90}{Packet drop  ratio (\%) }}} &
		\multicolumn{1}{c|}{\textbf{\rotatebox[origin=c]{90}{Precision (val)}}} & \multicolumn{1}{c|}{\textbf{\rotatebox[origin=c]{90}{Recall (val) }}} & \multicolumn{1}{c|}{\textbf{\rotatebox[origin=c]{90}{Accuracy (val)}}} & \multicolumn{1}{c|}{\textbf{\rotatebox[origin=c]{90}{Miss rate (val) }}} & \multicolumn{1}{c|}{\textbf{\rotatebox[origin=c]{90}{Detection rate or \textit{TPR} (val)}}} & 
		%

		\multicolumn{1}{c|}{\textbf{\rotatebox[origin=c]{90}{Affected child nodes (\%)}}}&
		\multicolumn{1}{c|}{\textbf{\rotatebox[origin=c]{90}{True positives (\%) }}} \\ \hline
		\endhead
		Dvir \textit{et al.}\cite{Dvir2011} & - & - & - & - & - & - & - & - & - & - & - & - & -& - &   - & - & - & - & -  & - & - & - & - & -  & - & -\\ \hline
		Landsmann \textit{et al.}\cite{landsmann2013topology} & - & - & - & - & - & - & - & - & - & - & - & - & - & - & - & - & - & - & -  & - &  - & - & - & -  & -& -\\ \hline
		Perrey \textit{et al.}\cite{PerreyLUSW13} &  &  20 & 2  & - & - & - &  -& - & - & - & - & - &-   & - &  - & - & - & - & -  & - & - & - & - & -  &- & -\\ \hline
		Seeber \textit{et al.}\cite{seeber2013towards} & - & - & - & - & - & - & - & - & - & - & - & - & - & - & - & - & - & - & - & - & - & - & - &  -  &- & -\\ \hline
		Sehgal \textit{et al.} \cite{Sehgal2014} & - &-  & - & 99 &  40 & 55 & - & - & - & - & - & - &-  &  -&  - & - & - & - & - & - &  - & -  & - &  - & - & -\\ \hline
		Mayzaud \textit{et al.}\cite{Mayzaud2015} & - & - &-  & 99  & 50 &50 & - & - & - &  -&-  & - &-  & - &-   & - &  -&  -& -  & - & - & - &   - &  - & - & -\\ \hline
		Ghaleb \textit{et al.}\cite{AddressingDAO} & - & - & - & 99 & - & - & 90 & 70 & 55  &  30 & - & - & - & - & - & - & - & - &  &  -  & - & - &   - & -  &- & -\\ \hline
		Iuchi \textit{et al.}\cite{iuchi2015secure} & - & -  & - & - & - &-  &  -& - & - &-  &-  & - &   -& -  &-  &   - &-  &-  &-   & - & - & - &   -  &  - &  99& -\\ \hline
		Glissa \textit{et al.}\cite{Glissa}&  -& - & - &93  & - & 35 &-  &  -& - & 35  & - & - &  -& - & - & - & - & - & - &  -  &  -& - &  -&-  &-  &- \\ \hline
		Djedjig \textit{et al.}\cite{Djedjig2017} & - &  - & - & - & - & - & - & - & - & - & - & - & - & - & - & - & - & - & - & - & - & - & - & -  & -  & -\\ \hline
		Ariehrour \textit{et al.}\cite{Airehrour2017AJTDE} & 63   &-  &-  &-  &-  & - & - & - &-  & - & 30 & 80 & - &  -  &   -& - &-  & - & -  & - &  -&  -  &  -& - & -& -\\ \hline
		Ariehrour \textit{et al.} \cite{Airehrour2017} & 66 & - & -  &-  &-  &-  & - & - & - &-  & - & 28 & 66 & - &  -  &   -& - &-  & - & - & - &  -&  -&-  &  - & - \\ \hline
		Ariehrour \textit{et al.}\cite{Airehrour2018}  & - &   -&-  & - &-  & - & - & - & - & - & - &  15& 62 &-  &-  & - & - & - & - & - &  -&  -& - & - & - & -  \\ \hline
		Nygaard \textit{et al.}\cite{nygaard2017intrusion} &-  &   -&  -&  -& 99 & - & - &  -&  -&-  & - & - & -  &    - &  0& - & - &-  &- & -  &  -& 100 & - & 100  & -& -\\ \hline
		Amin \textit{et al.}\cite{RIDES} & -  & - & - & - & - & - &-  &-  &-  & - & - &  -&-  &-  &-  & -  & 5 &  98& - &  -& - & - & - & 90  & -& -\\ \hline
		Le \textit{et al.}\cite{Le2011} & -  & - & - & - & - & - & - & - & - & - & - & - & - & - & - &  - & - & -  & - & - & - & - & - & -  & - &  -\\ \hline
		Raza \textit{et al.}\cite{RAZA20132661} & - &  -&-  & - & 99 & - & - &  -& - & 99 & - &  -& - & - & - &  -&-  &  - & - & - &  - & - &-  & 100  & - & -\\ \hline
		Kasinathan \textit{et al.}\cite{kasinathan2013denial} &-  &  - & - & - &  -& - & - & - & - & - &  -&-  & - &   -& - &-  & - & - &  -&-  &-  &- & - & -   & - & 100\\ \hline
		Kasinathan \textit{et al.}\cite{kasinathan2013ids} & - & - & - &  -&  -& - & - & - & - &-  &-  & - & - &-  &  - & - & - &  -& - & -  &  -&  - &-  &-    &- & - \\ \hline
		Zhang \textit{et al.}\cite{zhang2015intrusion}  & - &  - & - &  -&  -& - & - & - & - &-  &-  & - & - &-  &  - & - & - &  -& - & - &  -& - &-  &-    &- & - \\ \hline
		Pongle \textit{et al.}\cite{pongle2015real} & - &  - &-  &-  & 0 & 86 & - & - & - &  -&-   &  -& - & - &  -& - & -   & - & - &  -& - &-  & - & 94 & - & - \\ \hline
		Mayzaud \textit{et al.}\cite{Mayzaud2016Version} &  -&-  &  - & - & - & - &  -&  -& - &  -& - &  -& - &-  & -  & - & 0 & - &  -&  -& - & -  &  -& -   &  - & -\\ \hline
		Mayzaud \textit{et al.}\cite{Mayzaud2017}  &  -&- & - & - & - & - &  -&  -& - &  -& - &  -& - &-  & -  & - & 0 & - &  -&  -& - & -  &  -& -   &  - & -\\ \hline
		Surender \textit{et al.}\cite{Surendar2016} & 8 &  - &  -& 8 & 11 & 17 &  -&-  & - & - & - & - &  -&  -& - &-  &  -  & - & 38 & - & - &-  &  -& - &-  &  -\\ \hline
		Le \textit{et al.}\cite{Le2016} & - & -  & - & - & 0 & - & - &  -& - &-  &-  & - &-  & - & - & - & 0 &-  &-  & -   &  -&-  &  -&100   &-  &-   \\ \hline
		Lai \textit{et al.}\cite{lai2016detection} & - &- &  -&  -&-  & - &-  & - & - & - &  -& - &-  & - &-  &-  &  -  &  -& - &  100 & 100 & 100 & - & -  & - & -  \\ \hline
		Shreenivas \textit{et al.}\cite{Shreenivas2017}  &  -&-  &  - &-  & - &  -& - &-  & - & 0 &-  &  -& - & - &  - &   -& - &-  & - & - &  -& - &-  &100   & - &   - \\ \hline
		Chen \textit{et al.}\cite{chen2016defense}  & - &  - & - & - & - & - & - & - & - & - & - & - & - & - &-  & - & -  & -  & - & 100 & 100 &  100&0  & -  & - &  - \\ \hline
		Ahsan \textit{et al.}\cite{ahsan2017wormhole}   & - &  -& -&  -&  -&  -& - &  -&-  & 0 &  -& - & - &  -&-  &-  &   - & - & -  & - & - & - & - & 95  &-  & -  \\ \hline
		Gara \textit{et al.}\cite{gara2017intrusion} & - & -  & - &-  & - & 0 & - & - & - & - & - & - & - & - & - &-  &  -  & - & - & - & - & - & - & 100  & - & -   \\ \hline
		Gara \textit{et al.}\cite{GaraInderscience} & - & -  & - &-  & - & 0 & - & - & - & - & - & - & - & - & - &-  &  -  & - & -& - & - & - & - & 100  & - & -   \\ \hline
		Napiah \textit{et al.}\cite{Napiah2018} & - &    -& - & - & 0 &  -& - &-  & - & - & - & - &  -&-  &-  &-  & 0 & -&- & - &-  & 99 & - & 99 & - & -  \\ \hline
		Bostani \textit{et al.}\cite{Bostani2017} &  -&  -  &  -& - & - & - &  -& - & - &-  & - & - & - & - & - & - & 2 &   - & - & - & - & 97 & - & 96  & - &  - \\ \hline
		Shafique \textit{et al.}\cite{shafique2018detection} &-  & -&  -&  -& - &  -&-  &-  &  -& 0 & - &  -& - & 3  & 0 & 0 & - & - &  - &-  & - & 100 & - & - &-  & 99  \\ \hline
		Ioulianou \textit{et al.}\cite{ioulianou2018signature} &  - & - & - & - & - & - & - & - & - & - & - & - & - & - & - & - & - & - & - & - & - & - &  - & - & - &  - \\ \hline
		Kfoury \textit{et al.}\cite{kfoury2019self} & - &  - & - & - & - & - & - & - & - & - & - & - & - & - & - & - & - & - & - & -  & - & - &  - & - & - &  - \\ \hline
		Shukla \textit{et al.}\cite{shukla2017ml} & - &  - & - & - & - & - & - &  -& - &  -&-  & - & - & - &  -& - & - &-  &  - &  -& - & - & - & 93  &  -&-   \\ \hline
		
	\end{longtable}
\end{landscape}

\color{black}
\twocolumn
\section{Cross-layered security solutions for RPL}\label{crosslayered}
\textcolor{black}{RPL security is not restricted to network layer specific defense solutions. IEEE $ 802.15.4 $ MAC layer implements several features to provide security services such as confidentiality, integrity, and replay protection. Data confidentiality is achieved through symmetric key cryptography techniques based on Advanced Encryption Standard in Counter with CBC-MAC (AES-CCM) algorithm, message integrity through Message Authentication Code (MAC), and replay protection through monotonically increasing sequence numbers \cite{Perazzo2017,sastry2004security, amin2016comprehensive}. IEEE $ 802.15.4 $ MAC layer defines eight different security levels, which can be chosen as per the security requirements of the application. Oliveira \textit{et al.} \cite{oliveira2013NACF} proposed a network access control (NAC) security framework for 6LoWPAN networks. The proposed framework aims to control the access of nodes to the existing network using prior administrative authorization, and later applies security compliance on the authorized nodes for security management. The security mechanism of the framework is capable of defending the network from unknown attacks. The major limitations of the NAC security framework include the requirement of Lightweight Secure Neighbor Discovery for LLNs, secure reprogramming mechanism, and message authentication mechanism for implementing the proposed framework in a real network. The resource constrained nature of LLN nodes may limit some of these requirements. Moreover, the proposed framework is not implemented and analyzed for validation. Further, the authors extended their previous work \cite{oliveira2013NACF} and proposed a network admission control solution in \cite{oliveira2013NAC1,oliveira2016NAC2}. The proposed solution has three main tasks, i.e., node detection and authentication, node authorization, and data filtering. The main limitations of the proposed solution include: (1) inherits attacks from neighbor discovery and RPL protocols; (2) it uses symmetric encryption, which increases resource consumption of nodes. The authors suggested using data filtering on RPL control messages, and elliptic curve mechanisms for minimizing resource consumption of nodes.}

\section{Open issues, research challenges and future directions}\label{Open issues and research challenges}
In this section, we have discussed some open issues and research challenges that need to be studied and addressed. 

\textit{Security Against Newly Developed Routing Attacks}: One of the most concerning issues in IoT security is defense against newly developed attacks. DIO suppression \cite{Perazzo2017}, Routing choice intrusion \cite{zhang2015intrusion}, and ETX manipulation \cite{Shreenivas2017} are three such attacks which target the RPL network by degrading networks performance silently. Many other attacks specific to RPL are yet to be found and will require robust defense mechanisms. Very few efforts towards the development of defense mechanisms against such attacks have been carried out. Hence, several defense techniques for defending against newly discovered attacks need to be proposed. 

\textit{Scalability}: Most of the existing defense solutions have been tested on small network scenarios, but in the practical world, IoT is enabled by a large network of heterogeneous resource constrained nodes \cite{Dvir2011,khan2013wormhole, zhang2015intrusion,lai2016detection}. \textcolor{black}{The performance of existing solutions may degrade in the case of large network which puts IoT applications open to attackers.} In addition to it, most of the critical IoT applications require a minimum delay in information forwarding, hence the demand of fast-reacting and lightweight defense solutions is increasing in order to carry out seamless network operations. \textcolor{black}{These solutions must not degrade the QoS of the network while supporting high scalability.} Hence, research can be carried out towards the development of highly scalable lightweight defense solutions.

\textit{Mobility}: Lamaazi \textit{et al.} \cite{Lamaazi2016} showed that the performance of RPL is severely influenced by mobile nodes. The standard specification of RPL \cite{winter2012rpl} does not define any mechanism to support mobility. Thus, the overall network performance is degraded in the presence of mobile nodes. Some types of IoT nodes have dynamic characteristics (mobility), which lead to an increase in the number of link disconnections, collisions, and packet loss. \textcolor{black}{When these mobile nodes perform malicious activities, the network performance drastically degrades.} This leads to a rise in the number of problems that need to be addressed for securing RPL networks. In \cite{Aris2016, Medjek2015, Medjek2017} impact of the Version number and Sybil attack, respectively under mobility is analyzed. However, the impact of other attacks on RPL under mobility needs to be studied. \textcolor{black}{Most of the existing secure protocol and IDS based defense solutions for RPL consider the only static environment and may not be applicable for the mobile environment.}

\textit{Cryptography Challenges}: The key management is one of the significant challenges for resource constrained networks, which requires attention. Several defense solutions \cite{Dvir2011, landsmann2013topology,khan2013wormhole, Taylor, Glissa} use cryptography techniques like Hash Chain Authentication, Merkle Tree Authentication, and Dynamic Keying impose computational, memory, and energy overhead on resource constrained devices. These overheads affect node lifetime, which is an essential criterion for critical IoT applications, e.g., industrial, forest, and landslide monitoring. The development of lightweight cryptography based security solutions for RPL that are suitable for resource constrained devices is still a big challenge and needs to be addressed.

\textit{Resource Limitations for Machine Learning}: Utilization of ML for the development of RPL specific security solutions is still a big task because of resource constraints. ML is proven to be effective in securing various wireless and wired networks with abundant resources. Thus, the customization of ML algorithms needs to be done in order to be used in resource constrained IoT. \textcolor{black}{The efforts to address this challenge will lead to the development of lightweight signature and anomaly based IDS solutions which may be very useful in providing quick detection and facilitation of fast mitigation procedures.}  

\textit{Issues with Trust Based Secure RPL Protocols}: Defense solutions proposed in \cite{Airehrour2017AJTDE, Airehrour2018} require every node in the network to operate in a promiscuous mode, in order to overhear neighbor packet transmissions. Such requirements make these solutions unsuitable for resource constrained IoT nodes. Thus, improvements in existing trust based solutions without relying on such strict requirements must be carried out. 

\textit{Hardware Security}: Node tampering is one of the widely used methods for compromising a node and reprogramming it to perform malicious activities \cite{Le2013} in the network. All the insider attacks are performed by compromising a legitimate node, which is already a part of the IoT network. \textcolor{black}{ An attacker can reprogram a node with malicious functions like decreasing rank and increasing rank. In addition, a node can be reprogrammed in such a way that it skips checking rank function.} Moreover, node tampering may lead to shared secret keys getting exposed. Thus, the development of tamper-proof node design is an open research area. \textcolor{black}{It may also affect many factors involved in IoT security, and most importantly in the prevention from insider attacks.} Some authors have suggested using TPM \cite{seeber2013towards, Djedjig2017} for securing IoT devices against insider attacks. However, TPM adds an extra cost to IoT networks and maybe infeasible for some IoT applications.

\textit{Network Security Monitoring over Encrypted Traffic}: \textcolor{black}{The rapid growth in encrypted traffic is creating challenges for security monitoring and intrusion detection. Encryption is being used by digital business organizations as a primary tool for securing information. Encryption not only brings security to businesses, but it also benefits the attacker to evade detection \cite{apthorpe2017spying}. IoT specific IDS solutions present in the literature are developed based on the assumption of non-encrypted traffic. However, in the present scenario, IoT applications are using encryption due to the availability of resource-rich hardware. Hence this issue needs to be considered while developing IDS for current IoT applications. Encrypted Traffic Analytics (ETA) is one of the possible solutions that can be studied to address this issue.}

\subsection{Potential Areas for Future Research} 
In addition to previously discussed issues and challenges, we list potential research areas for upcoming researchers in this field.

\textit{Moving target IPv$ 6 $ defense}: By continually changing the IPv6 address of a device, the attacks including eavesdropping, denial-of-service, or man-in-the-middle attack can be defended. Moving target IPv$ 6 $ defense mechanisms provide such capability to devices. Lightweight moving target based defense mechanisms for securing resource constrained devices against targeted attacks can be explored in-depth. Also, research on achieving resilience using temporary-private IPv$ 6 $ addresses \cite{Mavani2019} can be carried out. 

\textit{Collaborative IDS}: These types of IDS leverage collaboration among sensor nodes and $ 6 $BR for efficient and quick detection of attackers. Very few research works present in the literature that focuses on the development of collaborative IDS and can be explored further.

\textit{Defense against coordinated attacks}: In the present scenario, the attackers are now targeting IoT networks using coordinated attack strategy. These attacks severely degrade the network's performance without being detected. Popular IDS like SVELTE \cite{RAZA20132661} are vulnerable to coordinated attacks. Thus, an efficient attack detection and mitigation solution to defend RPL against coordinated routing attacks needs to be developed.

\textit{Active Learning}: Data insufficiency is of the significant problems for ML-based IDS. This problem can be solved by active learning, which optimizes the model learning during the training phase. This research area has recently gained the attention of security researchers. This needs a more in-depth study for leveraging its use in the development of IoT based IDSs.

\textit{Encrypted Traffic Analytics}:  ETA utilizes network traffic information that is independent of protocol details, e.g., lengths and arrival times of flows. These details can be used irrespective of encrypted and encrypted traffic for security monitoring of networks. ETA is an emerging topic in the field of network security and can be applied to IoT security as well.

\textit{Key management}: Most of the IoT applications involve unattended device operation in an untrusted environment, where nodes may quickly become the target of attackers. In the secure mode of RPL, the nodes are pre-loaded with security keys, which can be considered as a significant security vulnerability due to a single point of failure. The development of scalable and efficient key management mechanisms like generation, management, and storage are the growing research areas in RPL security. The exiting WSN based key management solutions present in the literature can be improved and applied in RPL.

\textit{Energy efficient cryptography}: Traditional cryptography algorithms are capable of achieving a higher level of security. However, these algorithms are computation-intensive. Hence, they consume many resources.  Such algorithms cannot be directly used in IoT applications because energy resources are limited. Thus, the development of energy-efficient cryptography algorithms to achieve the required level of security with minimum energy consumption is an essential concern for IoT security in the present scenario.

\textit{Security of IPv$ 6 $ over the TSCH mode of IEEE $ 802.15.4 $e ($ 6 $TiSCH) networks}: Recently, $ 6 $TiSCH protocol \cite{vilajosana2019ietf} has been standardized to attain low-power, scalable, and highly reliable operations in industrial applications. $ 6 $TiSCH uses time-slotted channel hopping (TSCH) MAC with IPv$ 6 $ addressing to achieve industrial-grade performance. It is integrated with $ 6 $LoWPAN, RPL, and CoAP protocols. One of the important considerations of $ 6 $TiSCH is the requirement of node-to-node synchronization to prevent synchronization loops in the network. The attacks particular to RPL may disrupt node-to-node synchronization, which decreases throughput and increases communication latency. The research on the security of RPL and $ 6 $TiSCH combination is still in its early stage and is a potential research area for security researchers.

\textit{Addressing RPL specific flooding attacks}: There is no efficient and suitable solution specially designed for defending flooding attack against RPL protocol \cite{Mayzaud2016Taxanomy}. To defend the DIS attack, RPL parameters can be used for setting safety thresholds in the RPL protocol. For example, DIS interval can be used to block the neighbors who are sending DIS messages very frequently, i.e., DIS messages are received before the expiry of DIS interval. Outlier Detection (OD) methods can be used to detect the neighbors (attacker) with abnormal behavior. DIS and DIO flooding attacks can be detected using OD based IDS. The main advantage of using OD is that these methods impose significantly less overhead on resource constrained nodes. 

\textit{Security solutions for dynamic networks}: To provide RPL with the ability to work efficiently in a dynamic network (i.e., mobility scenario), many enhancements have been proposed in the literature. Several RPL mobility enhancements are EMA-RPL, MoMoRo , mRPL, Co-RPL, and ME-RPL. Most of the existing RPL security solutions like SVELTE, SecRPL, SecTrust-RPL, and SRPL assume static network topology and may not be suitable for dynamic scenarios. However, at present, there are many use-cases in which RPL is deployed in dynamic networks. Thus, the existing solutions must be improved to make them suitable for dynamic networks. Also, this requirement must be fulfilled by the defense solutions which may be developed in the future.  

\textit{Fog computing for RPL security}: Resource constrained nature of LLNs limits the usage of existing state-of-the-art security mechanisms. However, in the present scenario, this limitation may be handled by currently emerging computing paradigms. One such emerging computing paradigm is Fog computing, which can be leveraged for securing IoT applications. To develop security solutions based on the combination of Edge, Fog Computing, RPL, and $ 6 $LoWPAN is a potential research area. The resource constrained nature of LLN nodes must also be taken care of beforehand as they demand low complexity authentication, and low message overhead based security solutions.       

\section{Conclusion}\label{Conclusion}
Self-organization, self-healing, global connectivity, resource constrained, and open nature characteristics of IoT make it the best choice for the development of applications that make human life easier. However, these characteristics also expose IoT to attackers targeting users' security and privacy. The network layer is one of the most favorite targets of attackers in the case of wireless networks, and because most of the IoT devices communicate using wireless medium IoT is more prone to attackers. 
To support efficient routing in LLNs, the RPL protocol has been standardized. RPL protocol is vulnerable to different attacks, which include attacks inherited from WSN and attacks specific to RPL. In this paper, we presented an exhaustive study on various attacks and defense solutions, in particular to the RPL protocol. First, we discussed a taxonomy of attacks on RPL in which attacks are classified based on their primary targets, including resources, topology, and traffic. Then, a taxonomy of different RPL specific defense solutions present in the literature is proposed. Various research challenges, open issues, and
future research directions observed from the literature survey
are also discussed. We observed that the research related to defense solutions specific to secure RPL protocol and RPL specific IDS methods is still in the early phase and requires more attention for providing full-fledged security to IoT applications. 

\section*{\textcolor{black}{Acknowledgment}}
\textcolor{black}{This research was supported by the Ministry of Human Resource Development, Government of India.}

\bibliographystyle{IEEEtran}
\bibliography{mybibfile}

\end{document}